\documentclass[11pt]{article}

\usepackage{amsmath, amsthm, amssymb}
\usepackage{zed-csp}
\usepackage{graphicx}
\usepackage{color}
\usepackage{EventB}
\usepackage{bigpage}

\newtheorem{thm}{Theorem}[section]
\newtheorem{lemma}[thm]{Lemma}
\newtheorem{cor}[thm]{Corollary}

\theoremstyle{definition}
\newtheorem{example}{Example}[section]

\newcommand{\Bmachine}{\;\textbf{machine}\;}
\newcommand{\Bvariables}{\;\textbf{variables}\;}

\newcommand{\Binvariant}{\;\textbf{invariant}\;}

\newcommand{\Bevents}{\;\textbf{events}\;}

\newcommand{\brefinedby}{\preccurlyeq}

\def\Bwhere{\;\mbox{\bf where}\;}
\def\Bwhen{\;\mbox{\bf when}\;}
\def\Bthen{\;\mbox{\bf then}\;}
\def\Bend{\;\mbox{\bf end}}
\def\Bany{\;\mbox{\bf any}\;}

\def\Bthen{\;\mbox{\bf then}\;}
\def\Brefines{\;\mbox{\bf refines}\;}
\def\Bstatus{\;\mbox{\bf status}}

\title{A Peered Bulletin Board for Robust Use in Verifiable Voting Systems}

\author{Chris Culnane and Steve Schneider \\ Department of Computing, University of Surrey}

\date{\today}

\parskip=\medskipamount

\begin{document}

\maketitle

\begin{abstract}
The Web Bulletin Board (WBB) is a key component of verifiable election systems.  It is used in the context of election verification to publish evidence of voting and tallying that voters and officials can check, and where challenges can be launched in the event of malfeasance.  In practice, the election authority has responsibility for implementing the web bulletin board correctly and reliably, and will wish to ensure that it behaves correctly even in the presence of failures and attacks.  To ensure robustness, an implementation will typically use a number of peers to be able to provide a correct service even when some peers go down or behave dishonestly.   In this paper we propose a new protocol to implement such a Web Bulletin Board, motivated by the needs of the vVote verifiable voting system.  Using a distributed algorithm increases the complexity of the protocol and requires careful reasoning in order to establish correctness.  Here we use the Event-B modelling and refinement approach to establish correctness of the peered design against an idealised specification of the bulletin board behaviour.  In particular we show that for $n$ peers, a threshold of $t > 2n/3$ peers behaving correctly is sufficient to ensure correct behaviour of the bulletin board distributed design.  The algorithm also behaves correctly even if honest or dishonest peers temporarily drop out of the protocol and then return.  The verification approach also establishes that the protocols used within the bulletin board do not interfere with each other.  This is the first time a peered web bulletin board suite of protocols has been formally verified.
\end{abstract}

\section{Introduction}

Verifiable voting systems such as Pr\^et \`a Voter \cite{chaum05:e-vote,RyanBHSX09}, Scantegrity \cite{SIITakPk}, Helios\cite{DBLP:conf/uss/Adida08}, Wombat \cite{wombat-evote}, STAR-Vote \cite{starvote-jets} and Civitas \cite{DBLP:conf/sp/ClarksonCM08}
typically have a requirement to publish information concerning votes cast and how they have been processed, in order to provide verifiability.  Voters and other external parties are able to check the published information and challenge the election if any cheating has occurred.  Such systems are generally described using a ``Bulletin Board'' for publication: a repository of the information collected throughout the election, made publicly available for inspection.  

There are certain (generally implicit) security assumptions on the bulletin board: that once items are on the bulletin board then they will not be removed, that the final information given at the end of the election is fixed and cannot be adjusted, and that it will provide the same view of that information to all parties.   For example, Adida's characterisation \cite{adida206:e-vote} states that ``Cryptographic voting protocols revolve around a central, digital bulletin board. As its name implies, the bulletin board is public and visible to all, via, for example, phone and web interfaces. All messages posted to the bulletin board are authenticated, and it is assumed that any data written to the bulletin board cannot be erased or tampered with.''   Alternatively a bulletin board has been described as a ``broadcast channel with memory'' \cite{peters05:e-vote,cramer97:e-vote,DBLP:conf/sp/KustersTV12}, with a Web Bulletin Board treated as a public broadcast channel.

Achieving these properties in an implementation is not so straightforward.  A current view is that ``\textit{we don't know how to build a secure bulletin board}'' \cite{wagnerEVT13}, and to date there is no generally available implementation of a secure bulletin board.  In practice bulletin boards are generally implemented by collecting election information as it progresses, and publishing the information via a website, as done for example by Helios, Wombat, and STAR-Vote, or making it available via a git repository as in the Norway 2013 e-voting trial\cite{norway13}.   However, these are not tamper proof, and information can be changed on them unless there are additional safeguards such as the cryptographic mechanisms based on hash chains proposed by Heather and Lundin \cite{DBLP:conf/ifip1-7/HeatherL08}.   The design of STAR-VOTE uses multiple peers to tolerate faulty or malicious components, and has the election authority sign the bulletin board contents, thus changes can occur only with the collusion of the electoral authority.

%Although verifiability means that accidental data corruption or deliberate tampering with the election can be detected, in practice an Election Authority needs to be confident that it can protect against such accidental or deliberate corruption of the published information. 
%
The bulletin board presented in this paper arises from the need to implement a bulletin board as part of the vVote system being developed for the Victorian State election 2014 \cite{EVOTE2012:VEC}.   The Victorian State election runs over a two week period of ``early voting'' before election day itself, and the bulletin board is required to publish its information daily during the election.   For robustness and trust the bulletin board will be comprised of a number of peers to receive items, provide receipts, and publish information. The rate at which votes may be received means that the peers cannot sustain the overhead of a consensus protocol every time an item is posted, so they each maintain a local copy of their view of the bulletin board, and agree on the bulletin board only when it is time to publish.  A further challenge is that the bulletin board may need to reject some items, for example audit of a ballot previously used to vote, or any vote on a ballot previously used or audited, so that incompatible posts are not published.   We achieve this requirement provided a threshold of the peers are honest and operational the bulletin board will behave correctly, even in the presence of individual peers going down, external attacks and a minority of dishonest peers.   

This paper presents a new bulletin board protocol designed to run with a network of peers and to operate correctly when a threshold of the peers are honest and operational.   We provide a formal model and verification of the protocol, using the  framework of Event-B \cite{DBLP:books/daglib/0024570}.  We verify the protocol in the context of a Dolev-Yao attacker \cite{DBLP:journals/tit/DolevY83}, who has control over the network and a minority of peers.  

The paper is structured as follows:  Section~\ref{sec:protocol} presents and motivates the protocol, Section~\ref{sec:eventb} introduces the Event-B framework and the refinement approach to modelling and verification, and the way it will be applied to the protocol and Sections~\ref{sec:ibb}--\ref{sec:BB4} provide the details of the four stages of modelling aspects of the protocol and the verification proofs in terms of simulation.  Section~\ref{sec:liveness} discusses sufficient conditions for liveness, and Section~\ref{sec:conc} concludes with a discussion of what has been achieved, its relationship to related work, and its context.

\section{A peered bulletin board protocol} \label{sec:protocol}

We present an implementation of a bulletin board that accepts items to be posted (if they do not clash with previous posts), issues receipts, and periodically publishes what it has received.  The bulletin board published for any particular period must include all items that had receipts issued during that period.   Robustness is achieved through the use of several peered servers which cooperate on accepting items, issuing receipts, and publishing the bulletin board.  They make use of a threshold signature scheme which allows a subset of the peers above a particular threshold to jointly generate signatures on data.   The peers collectively provide the bulletin board service as long as a threshold of them are honest, and as long as a threshold of them are involved in handling any item posted to the bulletin board.  Thus the implementation is correct in the presence of communication failures, unavailability or failure of peers, and also dishonesty of peers.  The threshold $t$ required to achieve this must be greater than two-thirds of the total number $n$ of peers:  $t > 2n/3$.   There is no single point of failure: the system can tolerate failure or non-participation of any component, as long as a threshold of peers remain operational at any stage.  It also allows for different threshold sets of peers to be operational at different times.  For example, a peer may be rebooted during the protocol, missing some item posts, and may then resume participation.

The key properties we require for this bulletin board are:
\begin{description}
\item[(bb.1)] only items that have been posted to the bulletin board may appear on it;
\item[(bb.2)] any item that has a receipt issued must appear on the published bulletin board;
\item[(bb.3)] two clashing items must not both appear on the bulletin board;
\item[(bb.4)] items cannot be removed from the bulletin board once they are published.
\end{description}
It follows from bb.2 and bb.3 that if two items clash then receipts must not be issued for both of them.

The bulletin board provides a protocol for the posting of an item and its acknowledgement with a receipt, and provides another two related protocols for the publishing of the bulleting board: an optimistic one, and a fallback.

\subsection{Posting and acknowledgement}
\begin{figure}[t]
 \def\svgwidth{0.8\linewidth}
\bigskip
\begin{center}
 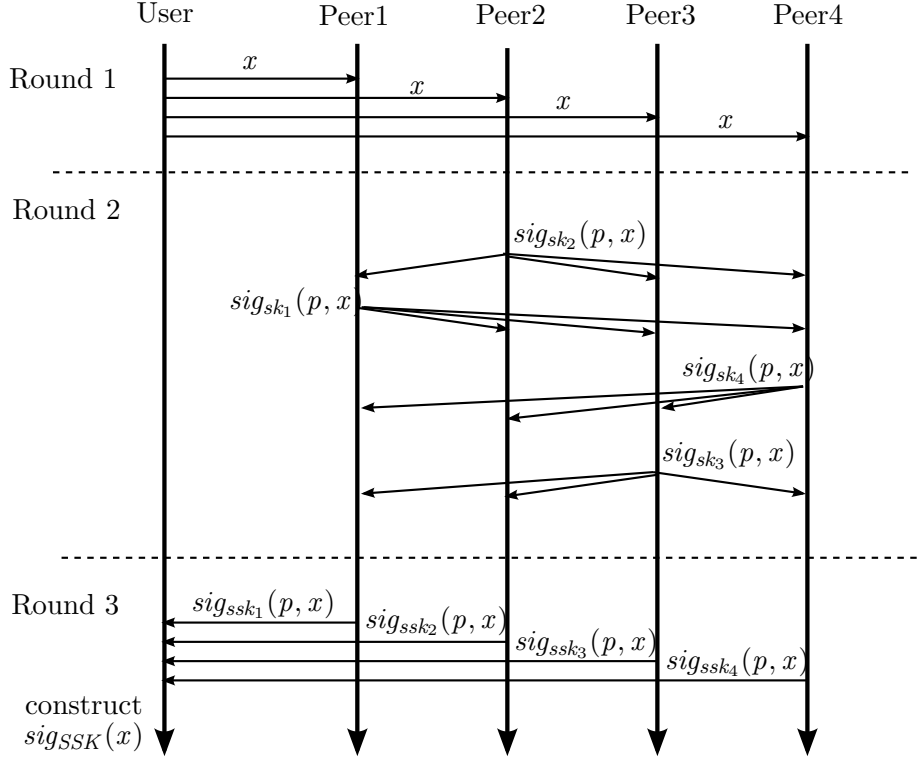
\end{center}
\caption{Posting Protocol}\label{fig:postingfull}
\end{figure}

The protocol for posting an item $x$ in period $p$, and issuing the acknowledgement, is as follows:
\[
\begin{array}{lllll}
1. & User \rightarrow P_i & : & x  & (\mbox{for each}\; i \in I) \\[0.5ex]
 \lefteqn{\qquad \quad \mbox{each $P_i$ checks no clash between $x$ and previous posts}} \\[0.5ex] 
2. & P_i \rightarrow P_j  & : & sig_{sk_i}(p,x) & (\mbox{for each}\;i,j \in I, j \neq i) \\[0.5ex]
 \lefteqn{\qquad \quad \mbox{each $P_i$ waits for at least a threshold number of signatures}} \\[0.5ex] 
3. & P_i \rightarrow User & : & sig_{ssk_i}(p,x) \qquad & (\mbox{for each}\;i \in I) 
\end{array}
\]
To post an item $x$, the User should first send $x$ to each of the peers, as shown in Round 1.  Each peer checks that $x$ does not clash with any posts it has received previously (from the current period or previous periods).  The peers then sign $(p,x)$ with their own individual signing key, and send the result to each of the other peers, as shown in Round 2.   Peers store all of the received signatures into their local database.  Finally, once a peer has obtained a threshold number of signatures on $(p,x)$ (including its own), it sends its share of the threshold signature on $(p,x)$ back to the User.  Once the User has received a threshold number of such shares it is able to combine them to provide a signature on $(p,x)$, and this serves as the receipt.  This protocol is shown in Figure~\ref{fig:postingfull}.  It is repeated for each item to be posted in the period.  

\subsection{Publishing the Bulletin Board}

The bulletin board is published at the end of the period.  The aim is for the peers to agree on the contents of the bulletin board and to issue their signature share on it to a public hosting service that can combine the signature shares and make the resulting signature publicly available.  

Peer $i$'s local record of the bulletin board $B_{i,p}$ is those items that it has received a threshold number of signatures on, which are those items it issues a signature share on towards the receipt. 

\begin{figure}[t]
 \def\svgwidth{0.8\linewidth}
\bigskip
\begin{center}
 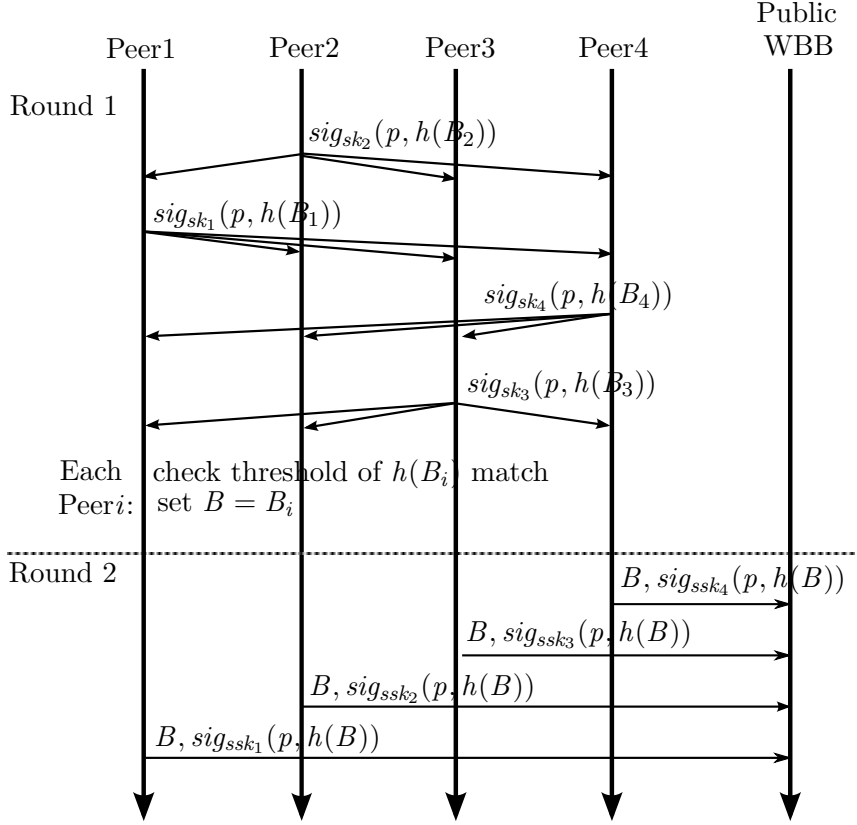
\end{center}
\caption{Optimistic Protocol}\label{fig:optimistic}
\end{figure}

The peers first of all run an optimistic protocol: this will succeed if at least a threshold of the local bulletin boards agree, which will be the case in practice if all peers are working properly.   The optmistic protocol is given as follows:
\[
\begin{array}{lllll}
1. & P_i \rightarrow P_j  & : & sig_{sk_i}(p,h(B_{i,p})) & (\mbox{for each}\;i,j \in I, j \neq i) \\[0.5ex]
 \lefteqn{\qquad \quad \mbox{each $P_i$ checks the hashes from a threshold of peers agree}} \\[0.5ex] 
2. & P_i \rightarrow WBB & : & B_{i,p},\, sig_{ssk_i}(p,h(B_{i,p})) \qquad & (\mbox{for each}\;i \in I) 
\end{array}
\]
The peers each sign a hash of their local copy of the bulletin board, and send them to each other.  If a threshold agree then they can issue the bulletin board and a share of the threshold signature on the hash.  This is illustrated in Figure~\ref{fig:optimistic}.

\begin{figure}[t]
 \def\svgwidth{0.8\linewidth}
\bigskip
\begin{center}
 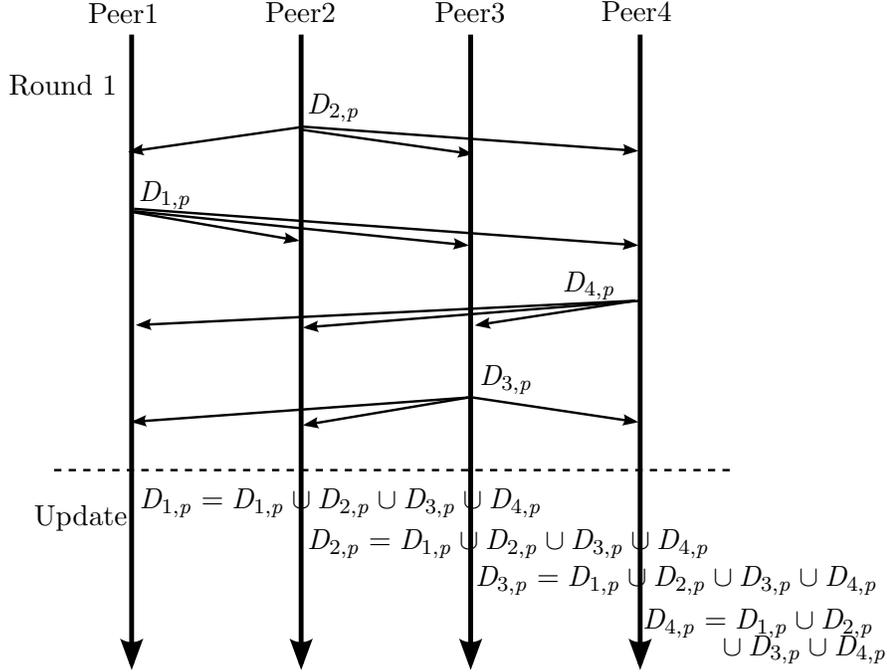
\end{center}
\caption{Fallback Protocol}\label{fig:fallback}
\end{figure}

If the optimistic protocol does not run successfully, because the hashes do not agree, that indicates that local bulletin boards are different.  In this case the peers exchange information about their bulletin boards using the fallback protocol as follows:
\[
\begin{array}{lllll}
1. & P_i \rightarrow P_j  & : & D_{i,p} \qquad & (\mbox{for each}\;i,j \in I, j \neq i) \\[0.5ex]
 \lefteqn{\qquad \quad \mbox{each $P_i$ adds any missing information received from others to its own database}}
\end{array}
\]
Each peer sends its database of signatures it has collected from the posting period to all the other peers, which update their databases with any signatures that are missing.  They can then recalculate their local bulletin board.  This is illustrated in Figure~\ref{fig:fallback}.  

After the fallback protocol is completed, the peers return to the optimistic protocol and repeat.  This is only required once for our liveness assumptions.   We assume for liveness either (1) that all peers are online and able to communicate during the fallback protocol (with no assumptions about the posting phase or correct behaviour of users), or (2) that a threshold of honest peers are online and able to communicate during the fallback protocol, and at every stage of the posting phase a threshold set of peers were live and able to communicate and that the posting users behaved honestly.   Under either of these two assumptions only one round of the fallback protocol is needed.  The difference with Byzantine Agreement protocols, which tend to require up to $(n-t)+1$ rounds to achieve agreement, is that the databases the peers start with have some consistency between them.  If thresholds of peers received posts correctly in the posting phase, then the honest peers involved in the exchange of information in the fallback protocol will all obtain the full bulletin board after one round.  Further explanation is provided in Section~\ref{sec:liveness}.

\section{Modelling and Verification Framework} \label{sec:eventb}

We use the action systems approach of Event-B~\cite{DBLP:books/daglib/0024570,rodin:d7} as our formal framework to model the protocol and to verify it.  In this approach systems are described in terms of the {\bf states} that they can be in, and the {\bf events} that transform the state.  

A system is defined as a {\em machine}, which encapsulates its state, and its events.  State information is described in terms of state variables and invariants on them.  The machine describes how the state is initialised, and how it can be updated with events.

The Event-B approach supports {\em refinement}, a relationship showing when one system implements another.   This approach allows a specification to be captured as an ideal machine that expresses the required behaviour.  An implementation satisfies the specification if it is a refinement.

Figure~\ref{fig:machine} illustrates how a machine is defined.  Machine $M$ is given with a list of state variables $v$, a state invariant $I(v)$, and a set of events $ev, \ldots$ to update the state.  Initialisation is a special event $init$ which sets the initial state of the machine, and its guard is $true$.

\begin{figure} 
\begin{center} 
\framebox{$\begin{array}{l}
\Bmachine \; M \\
\Bvariables \; v \\
\Binvariant \; I(v) \\
\Bevents \; init, ev, \ldots \\
\Bend \\ 
\end{array} \qquad 
\begin{array}{l}
ev \defs \\
\qquad \Bwhen  G (v) \\
\qquad \Bthen  v :| BA(v,v') \\
\qquad \Bend \\ 
\end{array}$}
\end{center}
\caption{Template of an Event-B machine and an event.}
\label{fig:machine}
\end{figure} 

Each event has a {\em guard} $G(v)$ over the variables $v$, and a {\em body}, usually written as an assignment $S$ on the variables.  The assignment is associated with a {\em before-after predicate} $BA(v, v')$ 
describing changes of variables upon event execution, in terms of the relationship between the variable values before ($v$) and after ($v'$).  For example, the assignment $v := v+1$ is associated with the predicate $v' = v+1$.  The body can also be written as $v :| BA(v,v')$, which assigns to $v$ any value $v'$ which makes the predicate $BA(v,v')$ true (see right of Fig.~\ref{fig:machine}, where $BA$ is the predicate in event $evt$).   In Event-B an event may also introduce local variables, which can be included in the guard (which constrains what values they can take), and in the body where they can be used to define the change of state.  Such events are constructed as: 
\[ \begin{array}{l}
evt \defs \\
\qquad \Bany  x \\
\qquad \Bwhere  G (v,x) \\
\qquad \Bthen  v :| BA(v,x,v') \\
\qquad \Bend \\ 
\end{array}
\]
Some of the conditions on $x$ may be included in the {\bf any} clause rather than the {\bf where} clause for readability (see e.g. $post$ and $a\_msg1$ of Figure~\ref{fig:bbu}).   Nondeterministic assignment has its own syntax: $x :\in S$ assigns $x$ some arbitrary element of $S$.  This is an abbreviation for \\
\begin{center}
{\bf any} $s$ {\bf where} $s \in S$ {\bf then} $x := s$ {\bf end}.
\end{center}

In this paper, all events have some feasible final state: whenever $G(v,x)$ is true then there is some $v'$ such that $BA(v,x,v')$ holds.

The Event-B approach to semantics, provided in \cite{DBLP:books/daglib/0024570,rodin:d7}, is to associate proof obligations with machines.  The key proof obligation on an event is that it preserves the invariant: when an event is called within its guard, then the state resulting from executing the body should meet the invariant.   For example, in the case of the machine in Fig.~\ref{fig:machine} we obtain the following proof obligation {\bf INV} on events which have the form of {\em evt}.  It states that if the invariant $I$ holds on $v$, and the guard $G(v)$ is true, and the before-after predicate relates $v'$ to $v$, then the invariant $I$ should be true on the state $v'$ reached after the event:
\begin{center}
$\Obligation
{
I (v) \land G(v) \land BA(v,v')
}
{
I(v') 
}
{INV}$ 
\end{center}
Discharging this proof obligation establishes that the event preserves the invariant.  The machine is consistent if this is true for all of its events.  It is true for all events in all machines presented in this paper: establishing this is one part of the proof of correctness.

\subsection{Event-B refinement} 

In Event-B, the intended refinement relationship between machines is directly written into the refinement machine definitions. As a consequence of writing a refining machine, a number of proof obligations arise.  Here, a machine and its refinement take the following form: \\

% \smallskip 
\begin{minipage}{0.45\textwidth}
$
\begin{array}{l}
\Bmachine \; M_0 \\
\Bvariables \; v \\
\Binvariant \; I(v) \\
\Bevents \; init_0, ev_0, ev'_0, \ldots \\
\Bend \\ \\ 
\end{array}
$
\end{minipage}
\begin{minipage}{0.45\textwidth}
$
\begin{array}{l}
\Bmachine \; M_1 \\
\Brefines \; M_0 \\
\Bvariables \; w \\
\Binvariant \; J(v,w) \\
\Bevents \; init_1, ev_1, ev'_1, \ldots \\
\Bend
\end{array}
$
\end{minipage} \\

%\medskip
\noindent The machine $M_0$ is refined by machine $M_1$, written $M_0 \brefinedby M_1$, if the given {\em linking invariant} $J(v,w)$ on the variables of the two machines is established by their initialisations, and  preserved by all events.  Any transition performed by a concrete event of $M_1$ can be matched by a step of the corresponding abstract event of $M_0$, or matched by $skip$ for newly introduced events, in order to maintain $J$.  This is similar to the approach of downwards simulation data refinement \cite{Derrick01a}, where the {\em simulation relation} plays the role of the linking invariant.  Formally, the refinement relation $M_0 \brefinedby M_1$ between abstract machine $M_0$ and concrete machine $M_1$ holds if the following proof obligations given below hold for all events:   

%\begin{figure}
%  \begin{center}
%    \framebox{$
%\begin{array}{l}
%ev_0 \defs \\
%\qquad \Bwhen G (v) \\
%\qquad \Bthen v :| BA0(v,v') \\
%\qquad \Bend \\ \\ 
%\end{array}
%\qquad 
%\begin{array}{l}
%ev_1  \defs \\
%\qquad \Brefines \; ev_0\\
%\qquad \Bwhen  H(w) \\
%\qquad \Bthen w :| BA1(w,w') \\
%\qquad \Bend
%\end{array}
%$
%}
%\end{center}
%\caption{An event and a refinement of it}
%\label{fig:events}
%\end{figure}
%
%\begin{figure}
%  \begin{center}
%    \framebox{$
%\begin{array}{l}
%ev_1 \defs \\
%\qquad \Bwhen  H (w) \\
%\qquad \Bthen  w :| BA1(w,w') \\
%\qquad \Bend 
%\end{array}
%$
%}
%\end{center}
%\caption{A new event not refining any event}
%\label{fig:events2}
%\end{figure}

\begin{description}

\item[GRD\_REF: Guard Strengthening]
%\subsubsection{Guard strengthening: GRD\_REF}

If a concrete event matches an abstract one, then this rule requires that when the concrete event is enabled, then so is the matching abstract one.  The rule is:
\begin{center}
$\Obligation
{
I (v) \land J (v, w) \land  H (w) 
}
{
G(v)
}
{GRD\_REF}$ 
\end{center}

\item[INV\_REF: Simulation]
%\subsubsection{Simulation: INV\_REF}
This ensures that the occurrence of events (including initialisation) in the concrete machine can be matched in the abstract one.  If there is a matching abstract event then the rule is:
\begin{center}
$\Obligation
{
I (v) \land J (v, w) \land H (w) \land BA1(w,w') 
}
{
\exists v' . (BA0(v,v') \land J(v',w'))
}
{INV\_REF$_1$}$ 
\end{center}
New events are treated as refinements of $skip$.  In this case the abstract state does not change (i.e., $v' = v$), and the rule is
\begin{center}
$\Obligation
{
I (v) \land J (v, w) \land H (w) \land BA1(w,w') 
}
{
J(v,w')
}
{INV\_REF$_2$}$ 
\end{center}
\end{description}

\subsubsection*{Refinement with respect to $A$}

It may be that an environment interacts with a machine $M_0$ only on some subset $A$ of its events.  In that case we can consider a refinement $M_1$ of $M_0$ with respect to $A$.  This requires that $M_1$ also has all the events $A$, and that GRD\_REF and INV\_REF$_1$ must hold for all the events in $A$.  However, other events of $M_1$ can be matched either by $skip$, or by some matching event (not in $A$) in $M_0$, in which case the guard must also match.  Thus for events not in $A$ we weaken the requirement to the single proof obligation GRD\_INV\_REF$_3$:
\begin{center}
$\Obligation
{
I (v) \land J (v, w) \land H (w) \land BA1(w,w') 
}
{
J(v,w') \lor (G(v) \land \exists v' . (BA0(v,v') \land J(v',w')))
}
{GRD\_INV\_REF$_3$}$ 
\end{center}
We will use this notion of refinement to express our requirements on the bulletin board protocol.

\subsection{Framework for Bulletin Board Modelling and Verification}

We are concerned with developing a peered bulletin board that can operate correctly in an unreliable environment, and  with some potentially misbehaving peers.  
In particular, communications between the bulletin board and its users may be under the control of an adversary, who may intercept, divert, block, duplicate and spoof messages.  The bulletin board is designed for use in in such an environment.

\begin{figure}[t]
 \def\svgwidth{0.4\linewidth}
\begin{center}
 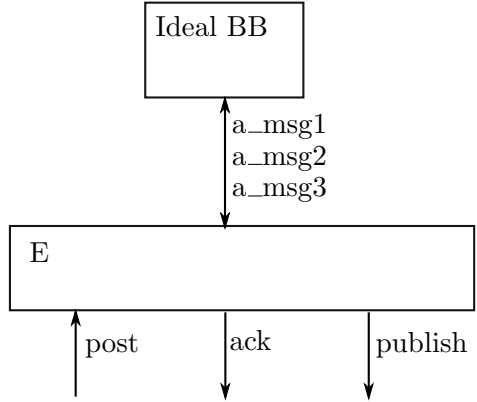
\end{center}
\caption{$BBSpec$:  ideal bulletin board and communication medium}\label{fig:abstract0}
\end{figure}

The specification of the bulletin board will encapsulate the required behaviour.  This will be described as an Event-B model $BBSpec$ with a description in terms of the architecture shown in Figure~\ref{fig:abstract0}, of an ideal bulletin board in the context of a reliable communication medium.   Users may use the events $post$, $ack$ and $publish$ to interact with the bulletin board, but communication occurs via the medium.  The bulletin board has its own corresponding interactions with the medium, labelled $a\_msg1$, $a\_msg2$ and $a\_msg3$.  These events are also within the model $BBSpec$, but they are not accessible directly to users.  Hence it is the behaviour of $BBSpec$ on the set of events $\{post, ack, publish \}$ that must be matched by any implementation.   

The bulletin board implementation uses a number of peers, for robustness and in order to distribute trust.   There are a total of $n$ peers, and we use a threshold signature scheme in which we require $t$ shares in order to produce a signature.   Our model of the protocol will be an Event-B model $BBProt$, in which we consider the adversary to control the communication medium to and from the peers and the $WBB$, and between them.  Hence any communication can be blocked.  We also consider that the adversary can control up to $n - t$ peers.   This means that such peers can sign and create any messages for sending, whether or not such messages are in accordance with the protocol, provided they have the appropriate keys.

We consider that (at least) a threshold $t$ of the $n$ peers are honest: that they follow the protocol.  Without loss of generality we will consider peers $1$ to $t$ to be honest, and $t+1$ to $n$ may behave arbitrarily (which includes honest behaviour).  This labelling of the peers captures the general case where some arbitrary $n-t$ peers may be dishonest, since the protocol is symmetric with respect to the labelling of the peers.  
% Recall that $t > 2n/3$, and so $n-t < n/3$.

\begin{figure}[t]
\def\svgwidth{0.5\linewidth}
\begin{center}
 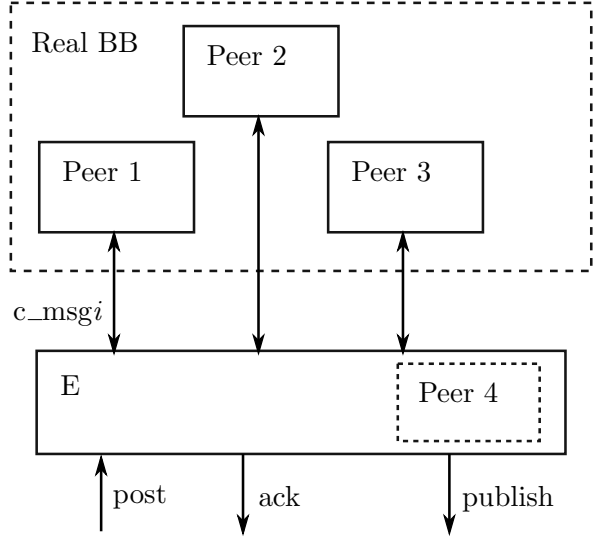
\end{center}
\caption{$BBProt$: Protocol model for analysis, with $t = 3$ and $n = 4$}\label{fig:concrete2}
\end{figure}

The model $BBProt$ includes the Dolev-Yao adversary, and peers $t+1$ to $n$ considered to be under the control of the adversary.    The setup is illustrated in Figure~\ref{fig:concrete2}.

$BBProt$ offers the same three external events as $BBSpec$, namely $post$, $ack$ and $publish$.  However it contains the peers explicitly, including peers controlled by the adversary, and so the communication patterns with and between the peers will be quite different to those in the specification.  Those communications are modelled by events $c\_msg i$.  The requirement for correctness will be that $BBSpec \brefinedby BBProt$ with respect to $\{post, ack, publish\}$.

The model $BBProt$ includes the events that make up the various bulletin board protocols.  Since these events can be performed whenever their guards are true, this means that interleavings of different protocols are naturally considered within this framework.  Thus our approach to verification automatically allows for possible interference between the protocols, and a proof of correctness establishes that the protocols cannot interfere in an adverse way.

\subsection{A useful lemma}

The following lemma and corollary will be useful in the refinement proofs.
\begin{lemma} \label{lem:counting}
If $A \subseteq \{ 1, \ldots, n \}$, $B \subseteq \{ 1, \ldots, n \}$, $\#A \ge t$, $\#B \ge t$, and $t > 2n/3$, then there is some $j \le t$ such that $j \in A$ and $j \in B$.
\end{lemma}
{\bf Proof}.  
We first prove that if $A \subseteq \{ 1, \ldots, n \}$, $B \subseteq \{ 1, \ldots, n \}$, $C \subseteq \{ 1, \ldots, n \}$, $\#A \ge t$, $\#B \ge t$, $\#C \ge t$, and $t > 2n/3$, then $\# (A \cap B \cap C) \ge 1$.

We use the law $\#X + \#Y = \#(X \cup Y) + \# (X \cap Y)$.  Observe that $A \cup B \subseteq \{ 1, \ldots, n\}$ and so $\#(A \cup B) \le n$.  We obtain:
\begin{eqnarray*}
\# (A \cap B) & = & \#A + \#B - \#(A \cup B) \\
& \ge & t + t - n
\end{eqnarray*}
Then
\begin{eqnarray*}
\# ((A \cap B) \cap C) & = & \#(A \cap B) + \#C - \#((A \cap B) \cup C) \\
& \ge & (2t - n) + t - n \\
& = & 3t - 2n
\end{eqnarray*}
Now $t > 2n/3$, so $3t - 2n > 0$.  Thus $\# ((A \cap B) \cap C) \ge 1$ as required.  

The result then follows immediately by setting $C = \{ 1, \ldots, t \}$: then there is some $j \in A \cap B \cap C$, i.e. $j \le t$ and $j \in A \inter B$.

\begin{cor} \label{cor:counting}
If $A \subseteq \{ 1, \ldots, t \}$, $B \subseteq \{ 1, \ldots, t \}$, $\#A \ge 2t-n$, $\#B \ge 2t-n$, and $t > 2n/3$, then there is some $j \le t$ such that $j \in A$ and $j \in B$.
\end{cor}
{\bf Proof}.  
The corollary follows from Lemma~\ref{lem:counting} on $A \union \{ t+1, \ldots, n \}$ and $B \union \{ t+1, \ldots, n \}$.

\section{One-shot Bulletin Board} \label{sec:ibb}

To structure the analysis, we will consider the protocol in four stages:
\begin{enumerate}
\item Firstly we have a single posting phase and a single commit phase for publication of the items posted to the bulletin board.
\item We next introduce multiple commit phases for multiple updates of the published bulletin board.
\item Next we allow for the BB to reject some posts based on previous posts
\item Finally we optimise each commit phase to optimistic and fallback, using hash functions. 
\end{enumerate}

Our first model, introduced here,  provides a one-shot bulletin board, which accepts posts for a period of time and then publishes its contents.  
\subsection{Specification}

We model the specified behaviour in terms of the bulletin board communicating with its environement over a medium as illustrated in Figure~\ref{fig:abstract0}.  We now give definitions for the events within that framework.   The given set $ITEM$ is the set of all items that can validly be posted to the bulletin board.  In practice there will be some mechanism for recognising a valid post, such as a signature, but for the purposes of this paper we abstract such a mechanism and assume that only elements of $ITEM$ are posted.  This corresponds to the expectation that posts not from $ITEM$ will be recognised and rejected by the bulletin board.

As described earlier, signatures are used to prevent the faking of receipts and the publishing of the bulletin board contents.  The bulletin board uses (threshold) signature key $SSK$ to sign receipts, and to sign the publication of the board.  We define 
\begin{eqnarray*}
RECEIPT & = & \{ sig_{SSK}(x) \mid x \in ITEM \} \\
PUBLISH & = & \{ sig_{SSK}(B) \mid B \subseteq ITEM \}
\end{eqnarray*}

\begin{figure}
\begin{center} 
\framebox{
$\begin{array}{l}
\Bmachine BBSpec1 \\
\Bvariables  E_A, R, C  \\ 
\Binvariant  E_A \subseteq ITEM \union RECEIPT \union PUBLISH \\
\phantom{\Binvariant} R \subseteq ITEM \\
\phantom{\Binvariant} C \subseteq ITEM \\
\Bevents  \\   
\quad \mbox{init} \defs E_A := \{ \} \parallel R := \{  \} \parallel C := \{ \}; \\[1ex]

\quad  \mbox{post}(x) \defs \Bwhen x \in ITEM \Bthen E_A := E_A \union \{ x \} \Bend ; \\[1ex]

\quad r \longleftarrow \mbox{ack} \defs r :\in (E_A \inter RECEIPT); \\[1ex]

\quad  P \longleftarrow \mbox{publish} \defs P :\in (E_A \inter PUBLISH); \\[1ex]

\quad  \mbox{a\_msg1} \defs \\
\qquad \Bany x \in E_A \inter ITEM \\
\qquad \Bthen R := R \union \{ x \}  \\
\qquad \Bend; \\[1ex]

\quad \mbox{a\_msg2} \defs \\
\qquad \Bany x \\
\qquad \Bwhere x \in R \land (sig_{SSK}(B) \in E_A \implies x \in B)  \\
\qquad \Bthen E_A := E_A \union \{ sig_{SSK}(x) \} \parallel C := C \union \{ x \} \\
\qquad \Bend; \\[1ex]

\quad \mbox{a\_msg3} \defs \\
\qquad \Bany Y \\
\qquad \Bwhere C \subseteq Y \subseteq R \land E_A \inter PUBLISH = \{ \} \\
\qquad \Bthen E_A := E_A \union \{sig_{SSK}(Y) \} \\
\qquad \Bend \\[1ex]

\Bend
\end{array}$
}
\end{center}
\caption{Bulletin Board Model incorporating the environment}
\label{fig:bbu}
\end{figure}

Allowing for untrusted peers requires us to include some nondeterminism within the specification of the bulletin board,
to reflect (bb.1) and (bb.2) above.  In particular, dishonest peers and the untrusted medium can prevent receipts from being issued for some received posts, so the specification must allow for this possibility.  
The model of the bulletin board thus uses two databases: $R$ consisting of received posts, and $C$ consisting of confirmed posts---those which have been acknowledged with receipts.  

When the board $B$ is published, anything published must be in $R$ in accordance with (bb.1); and all confirmed posts $C$ must be published in accordance with (bb.2).  Thus we require $C \subseteq B \subseteq R$.  In other words, items that have been submitted to the bulletin board but not confirmed might or might not appear in $B$.  We retain a level of uncertainty over what is published, because this level of uncertainty is present in the implementation when some of the bulletin board peers are untrusted.    Furthermore, in the implementation the adversary can orchestrate further posts and receipts following publication of the bulletin board, so our specification must reflect this: additional posts can be accepted.   
Requirement (bb.2) states that given both a published bulletin board and a receipted item, that item must be on the bulletin board.   To remain consistent with this requirement, any receipts issued in $a\_msg2$ after bulletin board publication must be on any published bulletin board.  Observe that if every posting has a receipt, then the bulletin board will contain all posted items  ($C = R = B$).   

Observe that $a\_msg3$ allows no more than one bulletin board to be published, meeting requirement (bb.4): once published, the bulletin board is fixed .  

The resulting Event-B model is given in Figure~\ref{fig:bbu}.  This is the specification that we will show our design meets.
The model includes a bulletin board and its environment.  As well as the state of the bulletin board, we include the state $E_A$ of the environment, containing the communications that it is managing, because we will want to consider the bulletin board protocol design in a model including the Dolev-Yao adversary, which provides an asynchronous communication medium.

\subsection{Implementation: a Robust Bulletin Board Design} \label{sec:wbbdesign}

The aim of the implementation is that if a threshold of peers behave according to the protocol, then the implementation will behave as the bulletin board of Section~\ref{sec:ibb} above with receipts and publication commitment.  This allows for a minority of peers to fail, or to behave maliciously, without impacting on the overall behaviour of the bulletin board.  In fact as we shall see, as long as a post $x$ is handled by some threshold of peers then a receipt can be provided, and $x$ will appear on the public web bulletin board.  Different posts can be handled by different threshold sets, allowing for individual peers to drop out temporarily (e.g. from a temporary loss of communication).

There are $n$ peers, numbered $1$ to $n$.  Each peer $j$ has its own signing key $sk_j$.  There is also a threshold signing key $SSK$, and each peer $j$ has a share of it: $ssk_{j}$.  Any $t$ out of $n$ partial signatures $SSK$ on a value $m$ can be combined to the corresponding signature on $m$:  $sig_{SSK}(m)$.  The condition on the threshold $t$ is that $t > 2n/3$.

The design is a slight simplification of that given in \cite{sds}.
Each peer $j$ maintains its local database $D_j$, which initially contains no entries.   It also has a boolean variable $pub_j$ which is initially {\em false}.

The peers run two protocols.  The first is for accepting posts and providing acknowledgements, and the second is for publishing the bulletin board.
\subsection*{Post and Acknowledge Protocol}
\begin{figure}[t]
 \def\svgwidth{0.8\linewidth}
\bigskip
\begin{center}
 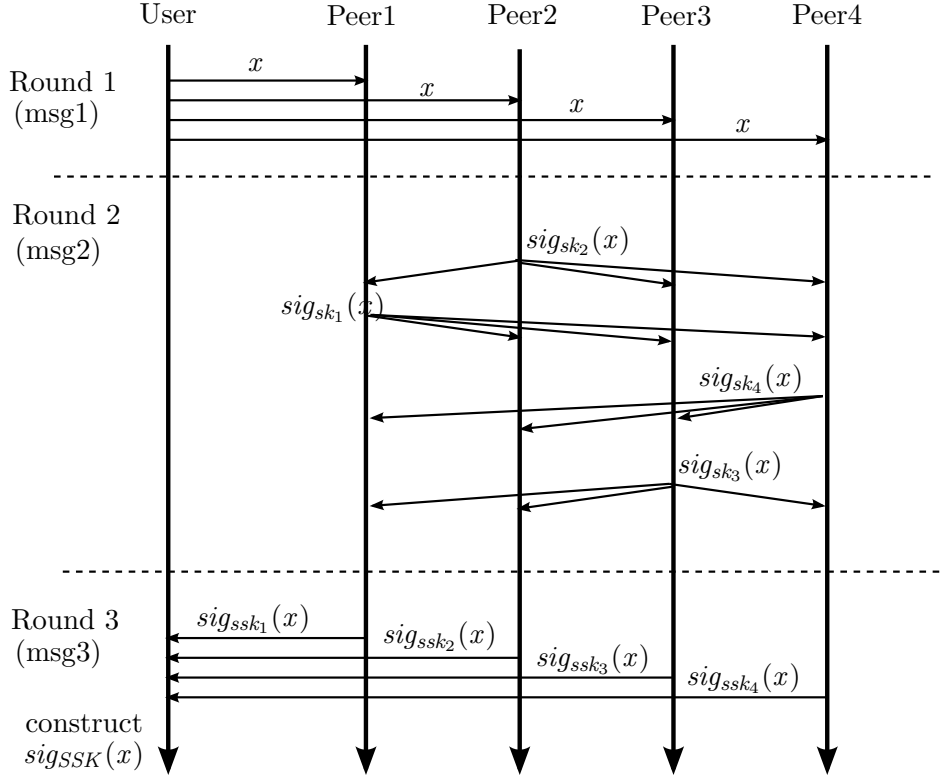
\end{center}
\caption{Posting Protocol}\label{fig:posting}
\end{figure}

This protocol is illustrated in Figure~\ref{fig:posting}, and is a simplication of the full protocol of Figure~\ref{fig:postingfull}.  It consists of three rounds, as follows:
\[
\begin{array}{lllll}
1. & User \rightarrow P_i & : & x  & (\mbox{for each}\; i \in I) \\[0.5ex]
2. & P_i \rightarrow P_j  & : & sig_{sk_i}(x) & (\mbox{for each}\;i,j \in I, j \neq i) \\[0.5ex]
 \lefteqn{\qquad \quad \mbox{each $P_i$ waits for at least a threshold number of signatures}} \\[0.5ex] 
3. & P_i \rightarrow User & : & sig_{ssk_i}(x) \qquad & (\mbox{for each}\;i \in I) 
\end{array}
\]

\subsection*{Publish Protocol} 

\begin{figure}[t]
 \def\svgwidth{0.8\linewidth}
\bigskip
\begin{center}
 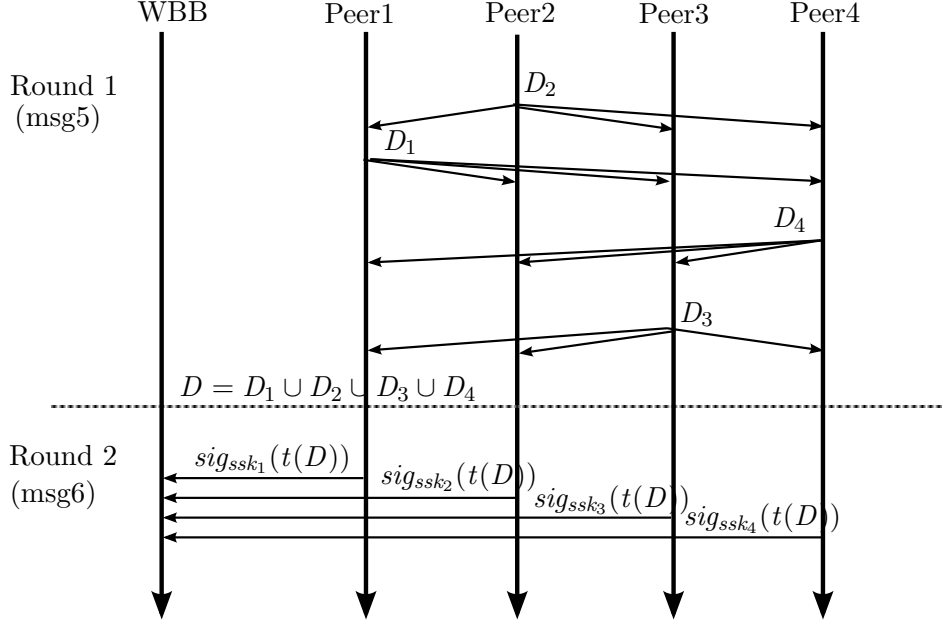
\end{center}
\caption{Publishing Protocol}\label{fig:publishing}
\end{figure}

This protocol is illustrated in Figure~\ref{fig:publishing}.  It is a combination of the pair of protocols given in Figures~\ref{fig:optimistic} and \ref{fig:fallback}, with Round 1 as the fallback protocol and then Round 2 as the optimistic protocol, and with the signature on the bulletin board directly.   When the time comes to publish, then the Post and Acknowledge protocol stops and is no longer executed, and the peer begins the commit protocol which is used for the peers to obtain agreement on the bulletin board to publish, as follows:
\[
\begin{array}{lllll}
1. & P_i \rightarrow P_j  & : & D_{i} \qquad & (\mbox{for each}\;i,j \in I, j \neq i) \\[0.5ex]
 \lefteqn{\qquad \quad \mbox{each $P_i$ adds any missing information received from others to its own database}} \\[0.5ex]
2. & P_i \rightarrow WBB & : & sig_{ssk_i}(t(D_i)) \qquad & (\mbox{for each}\;i \in I) 
\end{array}
\]

\subsection{Event-B Modelling and analysis}  \label{sec:modelling}

The threat model built into the model of the protocol incorporates our robustness considerations, in particular that the correctness of the bulletin board is not dependent on the correct behaviour of any individual component, as long as a threshold behave correctly.  It allows for the case where peers behave honestly but occasionally are down (either through connection loss, or through temporary server loss): this is modelled simply by the absence of messages between $DY$ and the peer, and allows for peers to miss some posts.  The model also includes the case where peers $t+1$ to $n$ can lose or otherwise alter their databases of received posts.  However, the honest peers $1$ to $t$ do not lose their databases: for correctness we require that a threshold of peers do not lose their data.   

The set $MESSAGE$ of all possible messages $m$ in the model is given as follows:
\begin{eqnarray*}
m & ::= & k \mid i \mid sig_k(m) \mid \{ m_1, \ldots, m_n \}
\end{eqnarray*}
where $k \in KEY$ and $i \in ITEM$.  Observe that a message can itself consist of a set of messages, and thus $MESSAGE$ covers all rounds of the protocol.  In particular $RECEIPT \subseteq MESSAGE$ and $PUBLISH \subseteq MESSAGE$.

Two further definitions will be useful when expressing the model:
\begin{eqnarray*}
SIG1 & = & \{ sig_{sk_k}(x) \mid 1 \le k \le n \land x \in ITEM \} \\
t(D) & = & \{ x \mid \# \{ k \mid sig_{sk_k}(x) \in D \} \ge t \} 
\end{eqnarray*}
The set $SIG1$ is the set of items signed by any of the peers.  Given a set $D$ of signed items, the set $t(D)$ is those items for which $D$ contains a threshold number of different signatures.  If $D$ is used to track the signed items received by a peer, then $t(D)$ is those items fpr which it has received a threshold number.

We will now define the model.  It is declared as follows:
\begin{center} 
\framebox{
$\begin{array}{l}
\Bmachine BBProt1 \\[1ex]
\Brefines BBSpec1 \\[1ex]
\Bvariables  E, I_j, D_j, pub_j, com_j \;\; (1 \le j \le t)
\end{array}$
}
\end{center}

Its invariant is given as follows:
\begin{center}
\framebox{
$\begin{array}{l}
\Binvariant  \\
\mbox{/* Types */} \\[0.5ex]
\qquad 
E \subseteq MESSAGE \\[0.5ex]
\qquad 
I_j \subseteq ITEM \\[0.5ex]
\qquad
D_j \subseteq SIG1 \\[0.5ex]
\qquad
pub_j \in BOOL \\[0.5ex]
\qquad
com_j \in BOOL \\[0.5ex]
\mbox{/* Key invariant properties */} \\[0.5ex]
\begin{minipage}{0.875 \textwidth}
\vspace*{-0.4cm}
\begin{eqnarray}
% && k \le t \land k \in c[x] \Rightarrow k \in d(x) \label{inv3} \\
&& k \le t \land k \in c[x] \land k \in s[B]  \Rightarrow  x \in B \hspace*{2in} \label{inv4} \\
&& sig_{ssk_j}(x) \in E  \implies  \# d_j[x] \ge t \label{invdj1} \\ 
&& sig_{SSK}(x) \in E  \implies  \# c[x] \ge t \label{inv4a} \\ 
&& sig_{ssk_j}(B) \in E \implies B \subseteq t(D_j) \label{invdj0} \\
&& sig_{SSK}(B) \in E  \implies  \# s[B] \ge t \label{inv4b} \\ 
&& D_j  \subseteq  E \label{invdj2} \\
&& k \le t \land k \in s[B] \implies com_k = true \label{com1} \\
&& k \le t \land k \in s[B_1] \land B_1 \neq B_2 \implies k \nin s[B_2]  \label{com2}
%&& sig_{sk_k}(m) \in E \land k \le t  \implies m \in ITEM \hspace*{2in} \label{inv6} \\
%&& sig_{ssk_k}(m) \in E \land k \le t  \implies m \in ITEM \lor m \subseteq ITEM 
 \end{eqnarray}
\mbox{/* adversary bound invariant --- see (\ref{dy0})--(\ref{dy2}) below} \\[1ex]
\mbox{/* Linking invariant --- see (\ref{link1})--(\ref{link3}) of Section~\ref{sec:sim} */ } \\[1ex]
where:\\
$\begin{array}[t]{rcll}
d_j[x] & = & \{ k \mid sig_{sk_{k}}(x) \in D_j \} & \mbox{shares of part sigs on $x$ received by Peer $j$}\\[0.5ex]
% d[x] & = & \{ k \mid sig_{SK_{1}}(x) \in R_k \} & \mbox{peers which hold a threshold signature on $x$} \\[0.5ex]
c[x] & = & \{ k \mid sig_{ssk_{k}}(x) \in E \} & \mbox{peers which have (part)signed the receipt on $x$} \\[0.5ex]
s[B] & = & \{ k \mid sig_{ssk_{k}}(B) \in E \} & \mbox{peers which have part-signed bulletin board $B$} 
\end{array}
$
\end{minipage}
\end{array}$
}
\end{center}
Each event introduced below preserves the invariant:  {\bf INV} is established for each event.

The initialisation and external events are given as follows:
\begin{center} 
\framebox{
$\begin{array}{l}
\Bevents  \\   
\quad \mbox{init} \defs E := \{ sk_{k} \mid k > t \} \union \{ ssk_k \mid k > t \} \parallel \\
\phantom{\quad \mbox{init} \defs} \Parallel_j (I_j := \{ \} \parallel D_j := \{ \} \parallel  pub_j := false \parallel com_j := false); \\[2ex]

\quad  \mbox{post}(x) \defs \Bwhen x \in ITEM \Bthen E := E \union \{ x \} \Bend ; \\[1ex]

\quad r \longleftarrow \mbox{ack} \defs r :\in (E_A \inter RECEIPT); \\[1ex]

\quad  P \longleftarrow \mbox{publish} \defs P :\in (E_A \inter PUBLISH);
\end{array}$
}
\end{center}

The initial state of each $Peer\;j$ has $I_j = \emptyset$, $D_j = \emptyset$, $pub_j = false$ and $com_j = false$.
\subsubsection*{Post and Acknowledge}
Recall that Round 2 of the posting protocol involves each peer sending a message to all the other peers.  This is split into two events: $c\_msg2a$ for the sending of the message (to be held by the Dolev-Yao environment), and $c\_msg2b$ for peers receiving the message.  We model the sending of the message to all peers by sending the message to the environment, and then allowing all the other peers to receive it.  
\begin{description}
\item[c\_msg1:] $DY \rightarrow Peer\; j : x$ \\  If $\neg pub_j$, then $Peer\; j$ inputs the post $x$.  
$Peer\;j$ adds $x$ to its local database of received items $I_j$.
\begin{center}
\framebox{
$\begin{array}{l}
\mbox{c\_msg1$_j$} \defs \\
\quad \Bany x \in E \inter ITEM \land \neg pub_j \\
\quad \Bthen I_j := I_j \union \{ x \}  \\
\quad \Bend;
\end{array}
$}
\end{center}
\item[c\_msg2a:] $Peer\; j \rightarrow DY : sig_{sk_{j}}(x)$ \hspace*{1cm} [$Peer\;j$ to all other peers.] \\  If $\neg pub_j$, and $I_j$ contains $x$ then Peer $j$ creates $sig_{sk_{j}}(x)$ with its signature key, outputs it to $DY$ intended for the other peers, and adds it to its local database $D_j$. 
\begin{center}
\framebox{
$\begin{array}{l}
\mbox{c\_msg2a$_j$} \defs \\
\quad \Bany x \\
\quad \Bwhere x\in I_j \land \neg pub_j \\
\quad \Bthen E := E \union \{ sig_{sk_{j}}(x) \} \parallel D_j := D_j \union \{ sig_{sk_{j}}(x) \}  \\
\quad \Bend;
\end{array}
$}
\end{center}
\item[c\_msg2b:] $DY \rightarrow Peer\; j : sig_{sk_{k}}(x)$ \hspace*{1cm} [$Peer\;j$ inputting from other peers.]\\  
If $\neg pub_j$, and $Peer_j$ inputs $sig_{sk_{k}}(x)$, then $Peer\;j$ adds $sig_{sk_{k}}(x)$ to $D_j$.  
\begin{center}
\framebox{
$\begin{array}{l}
\mbox{c\_msg2b$_j$} \defs \\
\quad \Bany x, k \\
\quad \Bwhere sig_{sk_{k}}(x) \in E \land \neg pub_j \\
\quad \Bthen D_j := D_j \union \{ sig_{sk_{k}}(x) \} \\
\quad \Bend;\end{array}
$}
\end{center}
\item[c\_msg3:] $Peer\;j \rightarrow DY : sig_{ssk_{j}}(x)$ \hspace*{1cm} \\  If $\neg pub_j$, and $D_j$ contains $t$ different signatures on $x$, then $Peer\;j$ outputs a signature share $sig_{ssk_j}(x)$.   
\begin{center}
\framebox{
$\begin{array}{l}
\mbox{c\_msg3$_j$} \defs \\
\quad \Bany x \\
\quad \Bwhere x \in t(D_j) \\
\quad \Bthen E := E \union \{ sig_{ssk_{j}}(x) \} \\
\quad \Bend; \end{array}
$}
\end{center}
\end{description}
When $DY$ has a threshold number of signature shares $sig_{ssk_{j}}(x)$ on $x$, $DY$ can combine them to form the receipt $sig_{SSK}(x)$, and add this to $E$.  (See event $c\_dy2$ below.)

\subsubsection*{Commit and Publish}

The publish protocol starts by setting $pub_j$ to true:
\begin{description}
\item[c\_msg4:] $Peer\;j: commit_j$ 
\begin{center}
\framebox{
$\begin{array}{l}
\mbox{c\_msg4$_j$} \defs \\
\quad \Bwhen \neg pub_j \\
\quad \Bthen pub_j := true \\
\quad \Bend; \end{array}
$}
\end{center}
\end{description}
Round 1 of the protocol, each peer sending a signature share to all the others, is modelled by two events: sending, and receiving.
\begin{description}
\item[c\_msg5a:] $Peer\;j \rightarrow DY : B_j$ \hspace*{1cm} [$Peer\;j$ to all other peers.]\\  If $pub_j$ then $Peer\;j$ outputs $B_j$, its local database of signed items, intended for the other peers.  This communicates its local database to the other peers.
\begin{center}
\framebox{
$\begin{array}{l}
\mbox{c\_msg5a$_j$} \defs  \\
\quad \Bwhen pub_j \\
\quad \Bthen E := E \union \{ D_j \} \\
\quad \Bend;\end{array}
$}
\end{center}
\item[c\_msg5b:] $DY \rightarrow Peer\;j : D_k$ \hspace*{1cm} \\  If $pub_j$ then $Peer\;j$ inputs $D_k$, $k$'s local database $D_k$.  This is added to $D_j$: any signed posts $sig_{sk_k}(x)$ in $D_k$ that are not already in $D_j$ are added to $D_j$.
\begin{center}
\framebox{
$\begin{array}{l}
\mbox{c\_msg5b$_j$} \defs \\
\quad \Bany D \\
\quad \Bwhere D \in E \land D \subseteq SIG1 \land pub_j \\
\quad \Bthen D_j := D_j \union D \\
\quad \Bend;\end{array}
$}
\end{center}
\item[c\_msg6:] $Peer\;j \rightarrow DY : sig_{ssk_j}(t(D_j))$ \\  If $pub_j$ then $Peer\;j$ can send out a signature share on its current version of the bulletin board: those items for which it holds a threshold of signatures, $t(D_j)$.
\begin{center}
\framebox{
$\begin{array}{l}
\mbox{c\_msg6$_j$} \defs \\
\quad \Bwhen pub_j \land \neg com_j \\
\quad \Bthen E := E \union \{ sig_{ssk_j}(t(D_j)) \} \parallel com_j := true \\
\quad \Bend;
\end{array}
$}
\end{center}
\end{description}

\subsubsection*{The Dolev-Yao environment}

The Dolev-Yao environment is modelled through the use of the set $E$ to retain all messages that are sent and received by protocol parties.  The adversary is also able to generate new messages to introduce into protocol executions.  In particular, he can sign any message with any key that he possesses; he can combine shares of a signature into a threshold signature; he can extract the message from a signature; and he can add and remove messages from a set of messages.  These capabilities are captured in the following derivation rules, which show how a new message can be generated from a set of messages.
\begin{eqnarray*}
\{ k, m \} & \vdash & sig_k(m) \\
\#S \ge t \implies \quad \{ sig_{ssk_{k}}(m) \mid k \in S \} & \vdash & sig_{SSK}(m) \\
\{ sig_k(m) \} & \vdash & m \\
\{ m, B \} & \vdash & B \union \{ m \} \\
m \in B \implies \quad \{ B \} & \vdash & m
\end{eqnarray*}
We model adversary behaviour by including an event for each rule, allowing the adversary to introduce new events to the set $E$.
\begin{center} 
\framebox{
$\begin{array}{l}
\quad \mbox{c\_dy1} \defs \qquad \mbox{/* signing */}\\
\qquad \Bany m, s  \\
\qquad \Bwhere m \in E \land s \in E \\
\qquad \Bthen E := E \union \{ sig_{s}(m) \} \\
\qquad \Bend; \\[1ex]

\quad \mbox{c\_dy2} \defs \qquad \mbox{/* threshold signature on $m$ */}\\
\qquad \Bany S, m \\
\qquad \Bwhere \# S \ge t \land \{ sig_{ssk_{k}}(m) \mid k \in S \} \subseteq E \\
\qquad \Bthen E := E \union \{ sig_{SSK}(m) \} \\
\qquad \Bend; \\[1ex]

\quad \mbox{c\_dy3} \defs \qquad \mbox{/* extracting $m$ from signature */}\\
\qquad \Bany m, s \\
\qquad \Bwhere sig_s(m) \in E \\
\qquad \Bthen E := E \union \{ m \} \\
\qquad \Bend; \\[1ex]

\quad \mbox{c\_dy4} \defs \qquad \mbox{/* adding $m$ to $B$ */}\\
\qquad \Bany m, B \\
\qquad \Bwhere m \in E \land B \in E \\
\qquad \Bthen E := E \union \{ B \union \{ m \} \} \\
\qquad \Bend; \\[1ex]

\quad \mbox{c\_dy5} \defs \qquad \mbox{/* extracting $m$ from $B$ */}\\
\qquad \Bany m, B \\
\qquad \Bwhere B \in E \land m \in B \\
\qquad \Bthen E := E \union \{ m \} \\
\qquad \Bend; \\[1ex]

\end{array}$
}
\end{center}

Some additional clauses are necessary to introduce into the invariant as below, to capture the limits of what the adversary can introduce.  These are necessary for the refinement proof.
\begin{center}
\framebox{
\begin{minipage}{0.9\textwidth}
{\mbox{ /* adversary bound invariant */}} 
\begin{eqnarray}
E \inter (\{ sk_k \mid k \le t \} \union \{ ssk_k \mid k \le t \}) & = & \emptyset \label{dy0} \\[1ex]
\Union_{e \in E} items(e) & \subseteq & E  \label{dy1} \\
\Union_{e \in E} sigs(e) & \subseteq & E  \label{dy2}
\end{eqnarray}
\mbox{where} \\
\begin{eqnarray*}
items(x) & = & \{ x \} \\
items(sig_s(x)) & = & items(x) \\
items(B) & = & \Union_{b \in B} items(b) \\[1ex]
sigs(x) & = & \{ \} \\
sigs(sig_s(x)) & = & \{ sig_s(x) \} \union sigs(x) \\
sigs(B) & = & \Union_{b \in B} sigs(b) \\[1ex]
\end{eqnarray*}
\end{minipage}
}
\end{center}

\subsection{Simulation} \label{sec:sim}

We aim to establish that the concrete system $BBProt1$ refines the abstract system $BBSpec1$ with respect to the external events $\{post, ack, publish \}$.  

To establish refinement we show that any concrete move can be matched by an abstract move, or (for events other that $post$, $ack$ and $publish$) matched by $skip$.  To do this we need to identify the {\em linking invariant}, the relationship between the abstract and concrete states, and show that any concrete move from a concrete state is matched for any corresponding abstract state by some abstract move or $skip$.  

\subsection*{Linking invariant}

We thus have to identify when, in the concrete system, abstract events are considered to have occurred.
\begin{itemize}
\item
abstract $a\_msg1$ occurs when the bulletin board receives $x$.  In the concrete model this corresponds to $t$ peers having received $x$ and signed it.  Since there can be up to $n-t$ dishonest peers, this means $t - (n-t) = 2t-n$ honest peers having signed $x$.
\item
abstract $a\_msg2$ occurs when the bulletin board issues a signature on $x$.  This corresponds to the combining of $t$ returns of signature shares $sig_{ssk_{j}}(x)$. 
\item
abstract $a\_msg3$ occurs when a signed database $sig_{SSK}(t(D))$ is produced.  This corresponds to the combining of $t$ returns of signature shares $sig_{ssk_{j}}(t(D))$. 
\end{itemize}

The abstract state $s_A$ is the pair of databases $R$ and $C$, and medium $E_A$.

The concrete state $s_C$ is the set of databases $I_j$, $D_j$ and $pub_j$ for the peers, $E$ for the Dolev-Yao environment.

The linking invariant is given by the following predicate $J(s_A, s_C)$:
\begin{center}
\framebox{
\begin{minipage}{0.9\textwidth}
{\mbox{ /* linking invariant */}} 
\begin{eqnarray}
%R & = & \{ x \in ITEM \mid sig_{SK_1}(x) \in R_l \;\mbox{for some}\; l \} \\
R & = & \{ x \in ITEM \mid \# \{ k \mid 1 \le k \le t \land sig_{sk_{k}}(x) \in E \} \ge 2t-n \} \label{link1} \\
% && {} \union \{ x \in ITEM \mid sig_{SK_1}(x) \in R_l \;\mbox{for some}\; l \} \\
C & = & \{ x \in ITEM \mid sig_{SSK}(x) \in E \} \label{link2} \\
E_A & = & E \inter (ITEM \union RECEIPT \union PUBLISH) \label{link3}
\end{eqnarray}
\end{minipage}
}
\end{center}
$R$ is the set of items for which the adversary (and possibly other peers) can provide a threshold of $sig_{sk_j}(x)$, and so can include $x$ on the published bulletin board.   If at least $2t-n$ honest peers have signed $x$, then it is within the adversary's control to produce a further $n-t$ signatures, giving a threshold of signatures on $x$.  $C$ is the set of items for which the adversary has a receipt---evidence that sufficiently many peers have a threshold of $sig_{sk_j}(x)$ to ensure that it will appear on the published bulletin board.

We are now in a position to present the main result: that the concrete model behaves according to the abstract model.

\begin{lemma} \label{lem:ref1}
$BBSpec1 \brefinedby BBProt1$ with respect to $\{ post, ack, publish \}$.
%
%If $J(s_A, s_C)$, and $s_C \stackrel{m_C}{\longrightarrow} s'_C$ then either $J(s_A, s'_C)$ ($m_c$ is matched by $skip$), or $\exists m_A, s'_A$ such that $s_A \stackrel{m_A}{\longrightarrow} s'_A$ and $J(s'_A, s'_C)$ ($m_c$ is matched by $m_A$).
\end{lemma}
\noindent {\bf Proof}
Consider each event of $BBProt1$ in turn.  It is necessary to prove {\bf GRD\_INV\_REF$_3$} in each case.  In most cases the event is matched by $skip$ and we establish {\bf INV\_REF$_2$}, which is stronger.  

\noindent {\bf Case} $post$.  Matched by $post$ of $BBSpec1$: the update to $E$ is matched by the update to $E_A$, preserving the linking invariant.

\noindent {\bf Case} $ack$.  Matched by $ack$ of $BBSpec1$:  the concrete output of receipt $r$ is matched by the abstract output of receipt $r$, since if $r \in E$ then $r \in E_A$ by the linking invariant.

\noindent {\bf Case} $publish$.  Matched by $publish$ of $BBSpec1$: the concrete output of $M$ from $E$ is matched by the abstract output of $M$, since if $M \in E$ then $M \in E_A$ by the linking invariant.

\noindent {\bf Case} $c\_msg1$.  Matched by $skip$ 

\noindent {\bf Case} $c\_msg2a$.  
If $\# \{ k \mid 1 \le k \le t \land sig_{sk_{k}}(x) \in E \} = 2t-n-1$ and $\# \{ k \mid 1 \le k \le t \land sig_{sk_{k}}(x) \in E' \} = 2t-n$ then this event is matched by $m_A = a\_msg1$ with $x$.  Otherwise matched by $skip$.  

\noindent {\bf Case} $c\_msg2b$.  Matched by $skip$.  

\noindent {\bf Case} $c\_msg3$.  Matched by $skip$. 

\noindent {\bf Case} $c\_msg4$.  Matched by $skip$.

\noindent {\bf Case} $c\_msg5_a$.  Matched by $skip$.

\noindent {\bf Case} $c\_msg5_b$.  Matched by $skip$.

\noindent {\bf Case} $c\_msg6$.  Matched by $skip$.

\noindent {\bf Case} $c\_dy1$.  Matched by $skip$.  In particular, $R$ remains unchanged due to invariant~(\ref{dy0}).

\noindent {\bf Case} $c\_dy2$.  For variable $E$, we use $E$ to refer to its value before the occurrence of this event, and $E'$ for its value after its occurrence.

If $x \in ITEM$ and $sig_{SSK}(x) \nin E$ and $sig_{SSK}(x) \in E'$, then this is matched by $a\_msg2$.  We must show that (1) $x \in R$ and (2) $sig_{SSK}(B) \in E_A \implies x \in B$.  
\begin{enumerate}
\item $x \in R$: We have that $\# c[x] \ge t$.  Hence there is some $k \le t$ with $k \in c[x]$, so by invariant~(\ref{invdj1}) it follows that $\# d_k[x] \ge t$.  By invariant~(\ref{invdj2}) it follows that $\# \{k \mid sig_{sk_{k}}(x) \in E \} \ge t$, and hence that $\# (\{k \mid sig_{sk_{k}}(x) \in E \}  - \{ t+1 \ldots n \}) \ge t - (n-t) = 2t-n$.  Hence $x \in R$ as required.
\item $sig_{SSK}(B) \in E_A \implies x \in B$:  Assume $sig_{SSK}(B) \in E_A$.  Then $\#s[B] \ge t$.  Also we have $\#c(x) \ge t$, so by  Lemma~\ref{lem:counting} there is some $k \le t$ with $k \in c[x]$ and $k \in s[B]$.  Hence from (\ref{inv4}) it follows that $x\in B$ as required.
\end{enumerate}

If $B_C \subseteq ITEM$ and $sig_{SSK}(B_C) \nin E$ and $sig_{SSK}(B_C) \in E'$, then this is matched by $a\_msg3$, with $B = B_C $.  
%The argument that $C \subseteq B \subseteq R$ follows the same pattern as the case for $c\_msg6$ above.
We must show that  (1) $E_A \inter PUBLISH = \{ \}$, (2) $C \subseteq B_C$ and (3) $B_C \subseteq R$.
\begin{enumerate}
\item $E_A \inter PUBLISH = \{ \}$: we establish this by contradiction.  If $sig_{SSK}(B) \in E_A$ for some $B \neq B_C$, then $\#s[B] \ge t$ by (\ref{inv4b}).  Also we have $s[B_C] \ge t$ by the guard of $c\_dy2$.  Hence from Lemma~\ref{lem:counting} there is some $k \le t$ with $k \in s[B]$ and $k \in s[B_C]$, contradicting (\ref{com2}).
\item $C \subseteq B_C$:  consider some $x \in C$.  Then $sig_{SSK}(x) \in E$, so $\# c[x] \ge t$ by invariant~(\ref{inv4a}).  Further, $\# s[B_C] \ge t$ by invariant~(\ref{inv4b}).  Hence from Lemma~\ref{lem:counting} there is some $k \le t$ with $k \in c[x]$ and $k \in s[B_C]$.  Hence by invariant~(\ref{inv4}), $x \in B_C$, as required.
\item $B_C \subseteq R$: We have from invariant~(\ref{inv4a}) that  $sig_{ssk_{j}}(B_C) \in E$ for some $j \le t$.  Now consider $x \in B_C$.  Then $x \in t(D_j)$ by invariant~(\ref{invdj0}).  Hence $x \in t(E)$ by invariant~(\ref{invdj2}), and so $\# \{ k \mid 1 \le k \le n \land sk_k(x) \in E \} \ge t$ from the definition of $t(E)$, and hence $\# \{ k \mid 1 \le k \le t \land sk_k(x) \in E \} \ge 2t-n$.  Thus $x \in R$ as required.
\end{enumerate}

Otherwise $c\_dy2$ is matched by $skip$. 

\noindent {\bf Case} $c\_dy3$.   Matched by $skip$, since $E_A = E \inter (ITEM \union RECEIPT \union PUBLISH)$ does not change, by invariants~(\ref{dy1}) and (\ref{dy2}).

\noindent {\bf Case} $c\_dy4$.  Matched by $skip$.

\noindent {\bf Case} $c\_dy5$.   Matched by $skip$, since any items, receipts or publish messages in $B$ are already in $E$ by invariants (\ref{dy1}) and (\ref{dy2}).

This concludes the proof that $BBSpec1 \brefinedby BBProt1$ with respect to $\{post, ack, publish \}$, establishing the correctness of the bulletin board protocol $BBProt1$ against the specification $BBSpec1$.

\subsection{Example attacks on weaker versions}

\begin{example} \label{ex:noround2}
In order to see the necessity for the round of signed messages in the posting and acknowledgement protocol (message 2), we consider what can occur if this round is not included.  In particular, if peers simply receive posts and respond with their signature share towards the receipt, then an adversary can organise for a receipt to be provided for an item not on the bulletin board, as follows:
\def\svgwidth{0.9\linewidth}
\begin{center}
 %% Creator: Inkscape inkscape 0.48.2, www.inkscape.org
%% PDF/EPS/PS + LaTeX output extension by Johan Engelen, 2010
%% Accompanies image file 'noRound2.pdf' (pdf, eps, ps)
%%
%% To include the image in your LaTeX document, write
%%   \input{<filename>.pdf_tex}
%%  instead of
%%   \includegraphics{<filename>.pdf}
%% To scale the image, write
%%   \def\svgwidth{<desired width>}
%%   \input{<filename>.pdf_tex}
%%  instead of
%%   \includegraphics[width=<desired width>]{<filename>.pdf}
%%
%% Images with a different path to the parent latex file can
%% be accessed with the `import' package (which may need to be
%% installed) using
%%   \usepackage{import}
%% in the preamble, and then including the image with
%%   \import{<path to file>}{<filename>.pdf_tex}
%% Alternatively, one can specify
%%   \graphicspath{{<path to file>/}}
%% 
%% For more information, please see info/svg-inkscape on CTAN:
%%   http://tug.ctan.org/tex-archive/info/svg-inkscape
%%
\begingroup%
  \makeatletter%
  \providecommand\color[2][]{%
    \errmessage{(Inkscape) Color is used for the text in Inkscape, but the package 'color.sty' is not loaded}%
    \renewcommand\color[2][]{}%
  }%
  \providecommand\transparent[1]{%
    \errmessage{(Inkscape) Transparency is used (non-zero) for the text in Inkscape, but the package 'transparent.sty' is not loaded}%
    \renewcommand\transparent[1]{}%
  }%
  \providecommand\rotatebox[2]{#2}%
  \ifx\svgwidth\undefined%
    \setlength{\unitlength}{439.52502441bp}%
    \ifx\svgscale\undefined%
      \relax%
    \else%
      \setlength{\unitlength}{\unitlength * \real{\svgscale}}%
    \fi%
  \else%
    \setlength{\unitlength}{\svgwidth}%
  \fi%
  \global\let\svgwidth\undefined%
  \global\let\svgscale\undefined%
  \makeatother%
  \begin{picture}(1,0.67719191)%
    \put(0,0){\includegraphics[width=\unitlength]{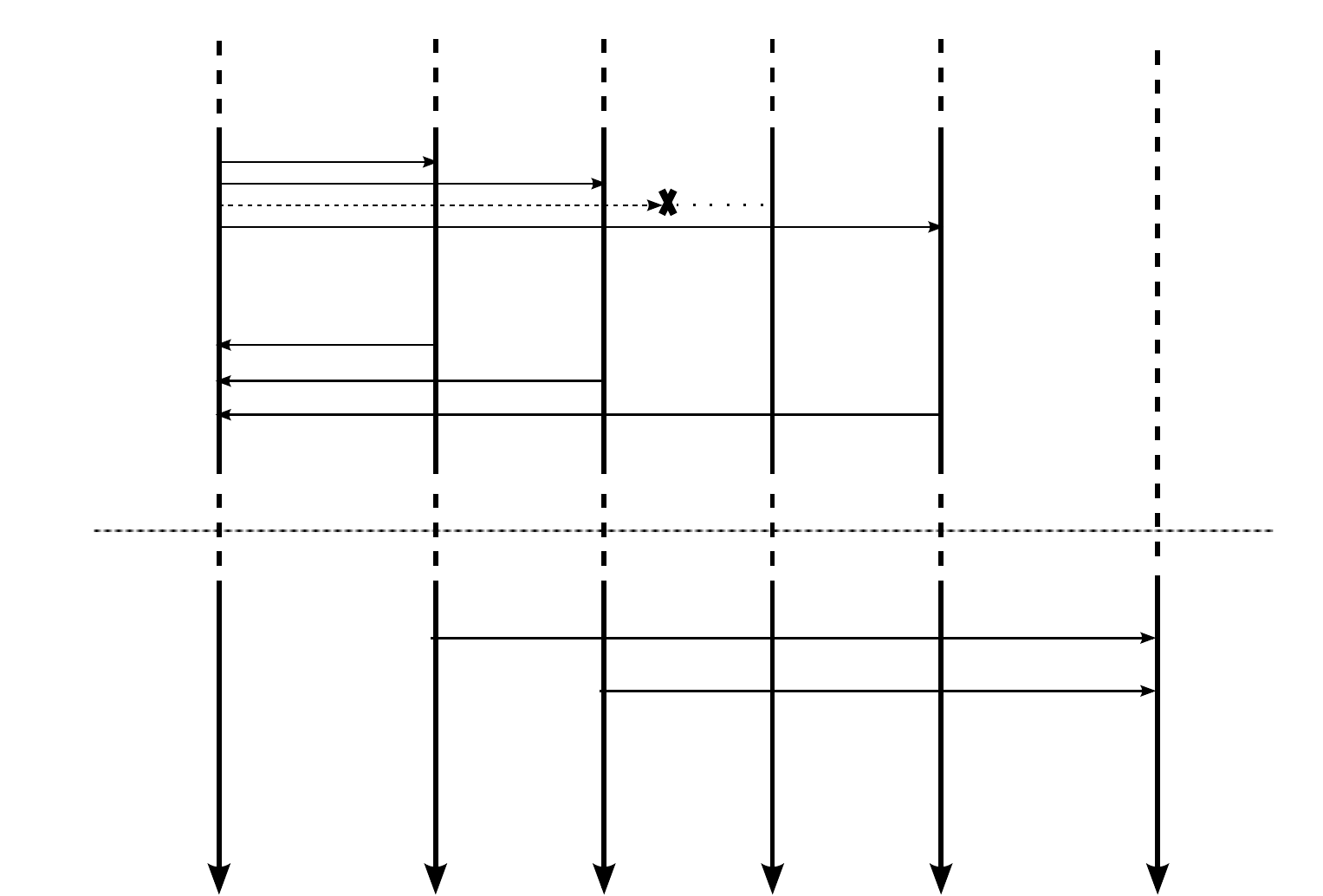}}%
    \put(0.2313831,0.56303052){\color[rgb]{0,0,0}\makebox(0,0)[lb]{\smash{$x$}}}%
    \put(0.1419359,0.66105843){\color[rgb]{0,0,0}\makebox(0,0)[lb]{\smash{User}}}%
    \put(0.29690838,0.65871824){\color[rgb]{0,0,0}\makebox(0,0)[lb]{\smash{Peer$1$}}}%
    \put(0.43003911,0.65949829){\color[rgb]{0,0,0}\makebox(0,0)[lb]{\smash{Peer$2$}}}%
    \put(0.55744938,0.65923828){\color[rgb]{0,0,0}\makebox(0,0)[lb]{\smash{Peer$3$}}}%
    \put(0.68251944,0.65975832){\color[rgb]{0,0,0}\makebox(0,0)[lb]{\smash{E:Peer$4$}}}%
    \put(0.18825517,0.42660026){\color[rgb]{0,0,0}\makebox(0,0)[lb]{\smash{$sig_{ssk_1}(x)$}}}%
    \put(0.34283763,0.40098823){\color[rgb]{0,0,0}\makebox(0,0)[lb]{\smash{$sig_{ssk_2}(x)$}}}%
    \put(0.59102763,0.37719629){\color[rgb]{0,0,0}\makebox(0,0)[lb]{\smash{$sig_{ssk_4}(x)$}}}%
    \put(0.38054066,0.54555988){\color[rgb]{0,0,0}\makebox(0,0)[lb]{\smash{$x$}}}%
    \put(0.63523118,0.51240719){\color[rgb]{0,0,0}\makebox(0,0)[lb]{\smash{$x$}}}%
    \put(0.02002162,0.33130257){\color[rgb]{0,0,0}\makebox(0,0)[lb]{\smash{$sig_{SSK}(x)$}}}%
    \put(0.02184177,0.36042492){\color[rgb]{0,0,0}\makebox(0,0)[lb]{\smash{construct}}}%
    \put(0.84766831,0.66126918){\color[rgb]{0,0,0}\makebox(0,0)[lb]{\smash{WBB}}}%
    \put(0.71765781,0.31050092){\color[rgb]{0,0,0}\makebox(0,0)[lb]{\smash{discard $x$}}}%
    \put(0.38561108,0.20519243){\color[rgb]{0,0,0}\makebox(0,0)[lb]{\smash{$x$ in some form}}}%
    \put(0.51744167,0.16306904){\color[rgb]{0,0,0}\makebox(0,0)[lb]{\smash{$x$ in some form}}}%
    \put(0.89395202,0.1453876){\color[rgb]{0,0,0}\makebox(0,0)[lb]{\smash{not
enough}}}%
    \put(0.02600211,0.54738002){\color[rgb]{0,0,0}\makebox(0,0)[lb]{\smash{Post $x$}}}%
    \put(0,0.20233217){\color[rgb]{0,0,0}\makebox(0,0)[lb]{\smash{\begin{tabular}[t]{l}
Construct\\
Bulletin\\ 
Board
\end{tabular}}}}%
    \put(0.71882178,0.34185843){\color[rgb]{0,0,0}\makebox(0,0)[lb]{\smash{Peer$4$}}}%
    \put(0.89398784,0.11393808){\color[rgb]{0,0,0}\makebox(0,0)[lb]{\smash{shares
of $x$}}}%
    \put(0.89310549,0.08136024){\color[rgb]{0,0,0}\makebox(0,0)[lb]{\smash{to include}}}%
    \put(0.89490719,0.05014057){\color[rgb]{0,0,0}\makebox(0,0)[lb]{\smash{on WBB}}}%
  \end{picture}%
\endgroup%

\end{center}
\end{example}
Here Peer 3 is cut out of the posting and acknowledgement protocol on a submission $x$, but a threshold of peers provide a signature share and so a receipt is provided.  However, peer 4 is dishonest, and so discards $x$ before publication of the bulletin board.   Hence there are only two shares of $x$ recorded, insufficient to warrant inclusion on the published bulletin board.

\begin{example} \label{ex:lowthreshold}
This example illustrates the necessity for the threshold to be greater than $2n/3$.  If the threshold is $2n/3$ or less, then an adversary can arrange for a receipt to be issued on an item not included on the bulletin board, as follows:

\def\svgwidth{0.8\linewidth}
\begin{center}
 %% Creator: Inkscape inkscape 0.48.2, www.inkscape.org
%% PDF/EPS/PS + LaTeX output extension by Johan Engelen, 2010
%% Accompanies image file 'lowThreshold.pdf' (pdf, eps, ps)
%%
%% To include the image in your LaTeX document, write
%%   \input{<filename>.pdf_tex}
%%  instead of
%%   \includegraphics{<filename>.pdf}
%% To scale the image, write
%%   \def\svgwidth{<desired width>}
%%   \input{<filename>.pdf_tex}
%%  instead of
%%   \includegraphics[width=<desired width>]{<filename>.pdf}
%%
%% Images with a different path to the parent latex file can
%% be accessed with the `import' package (which may need to be
%% installed) using
%%   \usepackage{import}
%% in the preamble, and then including the image with
%%   \import{<path to file>}{<filename>.pdf_tex}
%% Alternatively, one can specify
%%   \graphicspath{{<path to file>/}}
%% 
%% For more information, please see info/svg-inkscape on CTAN:
%%   http://tug.ctan.org/tex-archive/info/svg-inkscape
%%
\begingroup%
  \makeatletter%
  \providecommand\color[2][]{%
    \errmessage{(Inkscape) Color is used for the text in Inkscape, but the package 'color.sty' is not loaded}%
    \renewcommand\color[2][]{}%
  }%
  \providecommand\transparent[1]{%
    \errmessage{(Inkscape) Transparency is used (non-zero) for the text in Inkscape, but the package 'transparent.sty' is not loaded}%
    \renewcommand\transparent[1]{}%
  }%
  \providecommand\rotatebox[2]{#2}%
  \ifx\svgwidth\undefined%
    \setlength{\unitlength}{399.56958008bp}%
    \ifx\svgscale\undefined%
      \relax%
    \else%
      \setlength{\unitlength}{\unitlength * \real{\svgscale}}%
    \fi%
  \else%
    \setlength{\unitlength}{\svgwidth}%
  \fi%
  \global\let\svgwidth\undefined%
  \global\let\svgscale\undefined%
  \makeatother%
  \begin{picture}(1,0.8650378)%
    \put(0,0){\includegraphics[width=\unitlength]{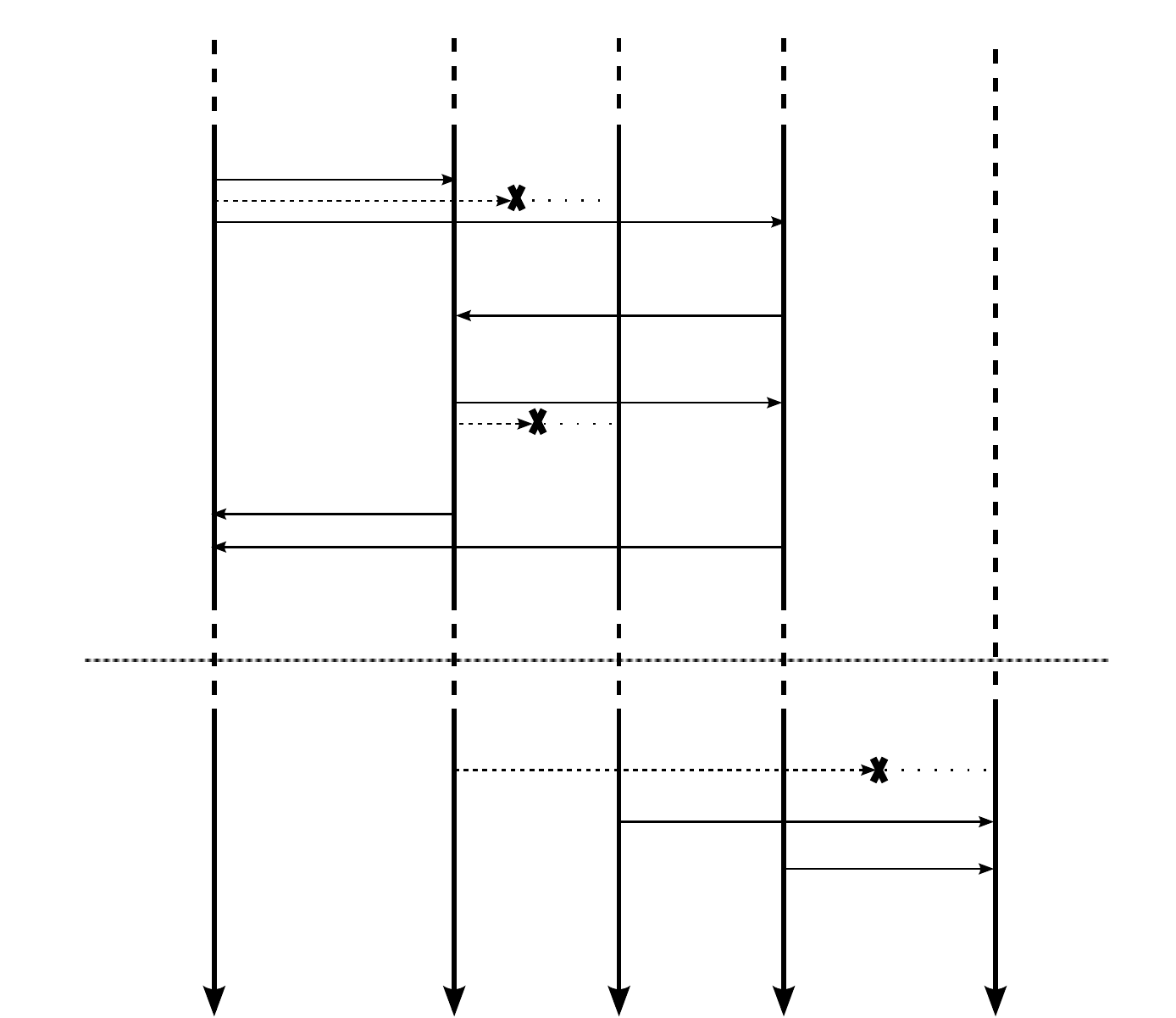}}%
    \put(0.15612894,0.84729103){\color[rgb]{0,0,0}\makebox(0,0)[lb]{\smash{User}}}%
    \put(0.35691644,0.84557489){\color[rgb]{0,0,0}\makebox(0,0)[lb]{\smash{Peer$1$}}}%
    \put(0.49706725,0.84528887){\color[rgb]{0,0,0}\makebox(0,0)[lb]{\smash{Peer$2$}}}%
    \put(0.63464385,0.84586092){\color[rgb]{0,0,0}\makebox(0,0)[lb]{\smash{E:Peer$3$}}}%
    \put(0.26099514,0.44108553){\color[rgb]{0,0,0}\makebox(0,0)[lb]{\smash{$sig_{ssk_1}(x)$}}}%
    \put(0.5340032,0.41491449){\color[rgb]{0,0,0}\makebox(0,0)[lb]{\smash{$sig_{ssk_3}(x)$}}}%
    \put(0.30246832,0.72024306){\color[rgb]{0,0,0}\makebox(0,0)[lb]{\smash{$x$}}}%
    \put(0.58262694,0.68377523){\color[rgb]{0,0,0}\makebox(0,0)[lb]{\smash{$x$}}}%
    \put(0.02202371,0.36443157){\color[rgb]{0,0,0}\makebox(0,0)[lb]{\smash{$sig_{SSK}(x)$}}}%
    \put(0.02402586,0.39646604){\color[rgb]{0,0,0}\makebox(0,0)[lb]{\smash{construct}}}%
    \put(0.81630698,0.84752286){\color[rgb]{0,0,0}\makebox(0,0)[lb]{\smash{WBB}}}%
    \put(0.67329591,0.34154984){\color[rgb]{0,0,0}\makebox(0,0)[lb]{\smash{discard $x$}}}%
    \put(0.39642657,0.22027646){\color[rgb]{0,0,0}\makebox(0,0)[lb]{\smash{$sig_{ssk_{1}}(t(D_1))$}}}%
    \put(0.02860222,0.72224522){\color[rgb]{0,0,0}\makebox(0,0)[lb]{\smash{Post $x$}}}%
    \put(-0,0.22256462){\color[rgb]{0,0,0}\makebox(0,0)[lb]{\smash{\begin{tabular}[t]{l}
Construct\\
Bulletin\\ 
Board
\end{tabular}}}}%
    \put(0.67457628,0.37604301){\color[rgb]{0,0,0}\makebox(0,0)[lb]{\smash{Peer$3$}}}%
    \put(0.53456014,0.17630006){\color[rgb]{0,0,0}\makebox(0,0)[lb]{\smash{$sig_{ssk_{2}}(t(D_2))$}}}%
    \put(0.67659878,0.13483918){\color[rgb]{0,0,0}\makebox(0,0)[lb]{\smash{$sig_{ssk_{3}}(t(D_2))$}}}%
    \put(0.85306027,0.09792892){\color[rgb]{0,0,0}\makebox(0,0)[lb]{\smash{$t(D_2)$ 
accepted}}}%
    \put(0.85259529,0.0633028){\color[rgb]{0,0,0}\makebox(0,0)[lb]{\smash{$x \nin R_2$}}}%
    \put(0.5392946,0.61412885){\color[rgb]{0,0,0}\makebox(0,0)[lb]{\smash{$sig_{sk_3}(x)$}}}%
    \put(0.39185024,0.53289859){\color[rgb]{0,0,0}\makebox(0,0)[lb]{\smash{$sig_{sk_1}(x)$}}}%
  \end{picture}%
\endgroup%

\end{center}
In this attack a receipt can be provided on a post $x$ although it does not appear on the board.   Peer 2 is excluded from the posting and acknowledgement of $x$, however participation from Peers 1 and 3 is sufficient to provide a receipt.  Peer 3 (which is dishonest) then discards $x$.  When the bulletin board is published, Peer 1 is excluded from the publication protocol, but Peers 2 and 3 agree on a bulletin board not including $x$, and so that is published.  The attack works because there is no honest peer that has participated in both the acknowledgement of $x$ and its posting on the bulletin board.  The attack cannot happen if the threshold is strictly greater than $2n/3$, because in that case there must be some honest peer contributing to both the receipt on $x$ and the agreed bulletin board, which is enough to ensure that $x$ is included on the bulletin board.
\end{example}

\section{Multiple Bulletin Board Rounds}
We extend to the case where multiple bulletin boards can be published.  We consider a period $p$ to consist of a number of posts followed by publication of the associated bulletin board for that period.  Thus different bulletin boards can be published for different periods, and we require that every period's bulletin board will behave according to the bulletin board specification given in $BBSpec1$. 

\subsection{Specification}

The specification of multiple bulletin boards is of a collection of boards that each behave according to specification $BBSpec1$.  This is captured as an indexed collection of bulletin boards within a single specification $BBSpec2$.
Receipts will be issued with the index of the bulletin board the item has been posted to, and a bulletin board will be published with its index.  We define 
\begin{eqnarray*}
RECEIPT2 & = & \{ sig_{SSK}(p,x) \mid x \in ITEM \land p \in \nat\} \\
PUBLISH2 & = & \{ sig_{SSK}(p,B) \mid B \subseteq ITEM \land p \in \nat \} \\
PUBLISH2_p & = & \{ sig_{SSK}(p,B) \mid B \subseteq ITEM  \}
\end{eqnarray*}
\begin{center} 
\framebox{
$\begin{array}{l}
\Bmachine BBSpec2 \\
\Bvariables  E_A, R_p, C_p \quad (p \in \nat)  \\ 
\Binvariant  E_A \subseteq ITEM \union RECEIPT2 \union PUBLISH2 \\
\phantom{\Binvariant}R_p \subseteq ITEM \quad (p \in \nat)  \\
\phantom{\Binvariant}C_p \subseteq ITEM \quad (p \in \nat)  \\
\Bevents  \\   
\quad \mbox{init} \defs E_A := \{ \} \parallel \Parallel_{p \in \nat} (R_p := \{  \} \parallel C_p := \{ \}) \\[1ex]

\quad  \mbox{post}(x) \defs \Bwhen x \in ITEM  \Bthen E_A := E_A \union \{ x \}  \Bend; \\[1ex]

\quad r \longleftarrow \mbox{ack} \defs r :\in (E_A \inter RECEIPT2); \\[1ex]

\quad  P \longleftarrow \mbox{publish} \defs P :\in (E_A \inter PUBLISH2); \\[1ex] 

\quad  \mbox{a\_msg1} \defs \\
\qquad \Bany x \in E_A \inter ITEM \land p \in \nat \\
\qquad \Bthen R_p := R_p \union \{ x \}  \\
\qquad \Bend; \\[1ex]

\quad \mbox{a\_msg2} \defs \\
\qquad \Bany x, p \\
\qquad \Bwhere x \in R_p \land (sig_{SSK}(p,B) \in E_A \implies x \in B) \\ 
\qquad \Bthen E_A := E_A \union \{ sig_{SSK}(p,x) \} \parallel C_p := C_p \union \{ x \} \\
\qquad \Bend; \\[1ex]

\quad \mbox{a\_msg3} \defs \\
\qquad \Bany Y, p \\
\qquad \Bwhere C_p \subseteq Y \subseteq R_p \land E_A \inter PUBLISH2_p = \{ \}  \\
\qquad \Bthen E_A := E_A \union \{sig_{SSK}(p,Y) \} \\
\qquad \Bend \\[1ex]

\Bend
\end{array}$
}
\end{center}

\subsection{Implementation}

In the implementation, each peer maintains a counter $p_j$ which it uses to track the period it is currently accepting posts for.  The counter will be incremented when it has finished accepting posts for one period and begins accepting posts for the next.  It also maintains a separate state space for each period.  For example, where $BBProt1$ used $R_j$ for $j$'s record of what it had received, $BBProt2$ will use $R_{j,p}$ for $j$'s record of what it received in period $p$, and so will have a separate set for each period.

The resulting model $BBProt2$ is given in the various clauses below.  The model is shown in the events within the description.   The key to the refinement proof is that the interleaving of the events across the different periods do not interfere, even though peers can progress their periods independently and can be involved in publication of one bulletin board while receiving items for another.

\subsubsection*{Declaration and Invariant}

$SIG1_p = \{ sig_{SSK}(p,x) \mid x \in ITEM \}$

\begin{center} 
\framebox{
$\begin{array}{llll}
\Bmachine BBProt2 \\[1ex]
\Brefines BBSpec2 \\[1ex]
\Bvariables  E,\; I_{j,p},\; D_{j,p},\;  p_j \;\; (1 \le j \le t) \\[2ex]
\Binvariant  \\
\mbox{/* Types */} \\[0.5ex]
\qquad E \subseteq MESSAGE \\[0.5ex]
\qquad I_{j,p} \subseteq ITEM \\[0.5ex]
\qquad D_{j,p} \subseteq \{ sig_{sk_{k}}(p,x) \mid x \in ITEM \} \\[0.5ex]
\qquad p_j \in \nat \\[0.5ex]
\mbox{/* Key invariant properties */} \\[0.5ex]
\qquad
 k \le t \land k \in c[p,x] \land k \in s[p,B]  \Rightarrow x \in B \hspace*{2.4in} \\[0.5ex]
\qquad
 sig_{ssk_j}(p,x) \in E  \implies  \# d_j[p,x] \ge t \\[0.5ex] 
\qquad
 sig_{SSK}(p,x) \in E  \implies  \# c[p,x] \ge t \\[0.5ex] 
\qquad
 sig_{ssk_j}(p,B) \in E \implies B \subseteq t(D_{j,p}) \\[0.5ex]
\qquad
 sig_{SSK}(p,B) \in E  \implies  \# s[p,B] \ge t \\[0.5ex] 
\qquad
 D_{j,p}  \subseteq  E \\[0.5ex]
\qquad
 k \le t \land k \in s[p,B]  \Rightarrow c_k > p \\[0.5ex]
\qquad
 k \le t \land k \in s[p,B_1] \land B_1 \neq B_2 \Rightarrow k \nin s[p,B_2] \\[0.5ex]
{\mbox{ /* linking invariant */}} \\[0.5ex]
\qquad
R_p  =  \{ x \in ITEM \mid \# \{ k \mid 1 \le k \le t \land sig_{sk_{k}}(p,x) \in E \} \ge 2t-n \}  \\[0.5ex]
\qquad
C_p  =  \{ x \in ITEM \mid sig_{SSK}(p,x) \in E \} \\[0.5ex]
\qquad
E_A =  E \inter (ITEM \union RECEIPT2 \union PUBLISH2) \\[0.5ex]
\mbox{where:} \\
\qquad
d_j[p,x] = \{ k \mid sig_{sk_{k}}(p,x) \in D_j \}  \qquad \mbox{shares of part sigs on $x$ received by Peer $j$}\\[0.5ex]
\qquad
c[p,x]  \hspace*{1.2mm} =  \{ k \mid sig_{ssk_{k}}(p,x) \in E \}  \qquad  \mbox{peers which have (part)signed the receipt on $x$} \\[0.5ex]
\qquad
s[p,B] \hspace*{0.5mm} =  \{ k \mid sig_{ssk_{k}}(p,B) \in E \} \qquad  \mbox{peers which have part-signed bulletin board $B$} 
\end{array}$
}
\end{center}

\subsubsection*{External events}

External events look very similar in $BBProt2$.  

\begin{center} 
\framebox{
$\begin{array}{l}
\Bevents  \\   
\quad \mbox{init} \defs E := \{ sk_{k} \mid k > t \} \union \{ ssk_k \mid k > t \}  \parallel \\
\phantom{\quad \mbox{init} \defs} \Parallel_{j,n} (I_{j,n} := \{ \} \parallel D_{j,n} := \{ \} \parallel  p_j := 0 \parallel c_j := 0); \\[2ex]

\quad  \mbox{post}(x) \defs \Bwhen x \in ITEM \Bthen E := E \union \{ x \}  \Bend; \\[1.5ex]

\quad r \longleftarrow \mbox{ack} \defs 
r :\in (E \inter RECEIPT2); \\[1ex]

\quad  P \longleftarrow \mbox{publish} \defs P :\in E \inter PUBLISH2; 
\end{array}$
}
\end{center}

\subsubsection*{Posting and acknowledgement protocol}
Posting and acknowledgement is similar.  The new aspect is the introduction of the period $p_j$, and $Peer_j$ may only accept and acknowledge items, and issue its share of the receipt, for items in its current period.

\begin{center} 
\framebox{
$\begin{array}{l}

\quad  \mbox{c\_msg1$_j$}(x) \defs \qquad \mbox{/* receive item $x$ */} \\
\qquad \Bwhen x \in E \inter ITEM \\
\qquad \Bthen I_{j,p_j} := I_{j,p_j} \union \{ x \}  \\
\qquad \Bend; \\[1ex]

\quad \mbox{c\_msg2a$_j$} \defs\qquad \mbox{/* send signature share on $x$ */} \\
\qquad \Bany x \\
\qquad \Bwhere x\in I_{j,p_j} \\
\qquad \Bthen E := E \union \{ sig_{sk_{j}}(p_j,x) \} \parallel D_{j,p_j} := D_{j,p_j} \union \{ sig_{sk_{j}}(p_j, x) \}  \\
\qquad \Bend; \\[1ex]

\quad \mbox{c\_msg2b$_j$} \defs \qquad \mbox{/* receive signature share on $x$ */} \\
\qquad \Bany x, k \\
\qquad \Bwhere sig_{sk_{k}}(p_j,x) \in E \\
\qquad \Bthen D_{j,p_j} := D_{j,p_j} \union \{ sig_{sk_{k}}(p_j,x) \} \\
\qquad \Bend; \\[1ex]

\quad \mbox{c\_msg3$_j$} \defs \qquad \mbox{/* send signature share on receipt of $x$ */} \\
\qquad \Bany x \\
\qquad \Bwhere x \in t(D_{j,p_j}) \\
\qquad \Bthen E := E \union \{ sig_{ssk_j}(p_j,x) \} \\
\qquad \Bend; 
\end{array}$
}
\end{center}

\subsubsection*{Commit and publish protocol}

The commit protocol for the bulletin board of period $p_j$ is started by incrementing $p_j$.  Thus no further posts will be accepted for that bulletin board, and the events in the commit and publish protocol are then enabled.  They match the events from $BBProt1$.  

\begin{center} 
\framebox{
$\begin{array}{l}
\quad \mbox{c\_msg4$_j$} \defs \qquad \mbox{/* start commit protocol */} \\
\qquad \Bbegin \\
\qquad p_j := p_j + 1 \\
\qquad \Bend; \\[1ex]

\quad \mbox{c\_msg5a$_j$} \defs \qquad \mbox{/* send database */} \\
\qquad \Bany p < p_j \\
\qquad \Bthen E := E \union \{ D_{j,p} \} \\
\qquad \Bend; \\[1ex]

\quad \mbox{c\_msg5b$_j$} \defs \qquad \mbox{/* receive $k$'s database, update $D_{j,p}$ if necesssary */} \\
\qquad \Bany D, p \\
\qquad \Bwhere D \in E \land D \subseteq SIG1_p \land p < p_j \\
\qquad \Bthen D_{j,p} := D_{j,p} \union D \\
\qquad \Bend; \\[1ex]

\quad \mbox{c\_msg6$_j$} \defs \qquad \mbox{/* publish signature share on $t(D_{j,p})$ */} \\
\qquad \Bwhen c_j < p_j \\
\qquad \Bthen E := E \union \{ sig_{ssk_j}(c_j,t(D_{j,c_j})) \} \parallel c_j := c_j + 1 \\
\qquad \Bend; \\[1ex]

\end{array}$
}
\end{center}

\subsubsection*{Dolev-Yao environment}
The adversary has the same moves as before, with two new ones, combining and separating pairs.  This arises from the introduction of pairing in this model, to allow the period along with the message to be signed.  

\begin{center} 
\framebox{
\begin{minipage}{0.5\textwidth}
$\begin{array}{l}
 \mbox{c\_dy1} \defs \quad \mbox{/* signature share on $m$ */}\\
\quad \Bany m, s \\
\quad \Bwhere m \in E \land s \in E \\
\quad \Bthen E := E \union \{ sig_s(m) \} \\
\quad \Bend; \\[1ex]

 \mbox{c\_dy2} \defs \quad \mbox{/* threshold signature on $m$ */}\\
\quad \Bany S, m \\
\quad \Bwhere \# S \ge t \land \{ sig_{ssk_{k}}(m) \mid k \in S \} \subseteq E \\
\quad \Bthen E := E \union \{ sig_{SSK}(m) \} \\
\quad \Bend; \\[1ex]

 \mbox{c\_dy3} \defs \quad \mbox{/* extracting $m$ from signature */}\\
\quad \Bany m, s \\
\quad \Bwhere sig_s(m) \in E \\
\quad \Bthen E := E \union \{ m \} \\
\quad \Bend; \\[15ex]
\end{array}$
\end{minipage}
\begin{minipage}{0.45\textwidth}
$\begin{array}{l}
 \mbox{c\_dy4} \defs \quad \mbox{/* adding $m$ to $B$ */}\\
\quad \Bany m, B \\
\quad \Bwhere m \in E \land B \in E \\
\quad \Bthen E := E \union \{ B \union \{ m \} \} \\
\quad \Bend; \\[1ex]
 \mbox{c\_dy5} \defs \quad \mbox{/* extracting $m$ from $B$ */}\\
\quad \Bany m, B \\
\quad \Bwhere B \in E \land m \in B \\
\quad \Bthen E := E \union \{ m \} \\
\quad \Bend; \\[1ex]

 \mbox{c\_dy6} \defs \quad \mbox{/* pairing */}\\
\quad \Bany m, p \\
\quad \Bwhere m \in E \land p \in \nat \\
\quad \Bthen E := E \union \{ (p,m) \} \\
\quad \Bend; \\[1ex]

 \mbox{c\_dy7} \defs \quad \mbox{/* splitting */}\\
\quad \Bany m, p \\
\quad \Bwhere (p,m) \in E \\
\quad \Bthen E := E \union \{ p, m \} \\
\quad \Bend 
\end{array}$
\end{minipage}
}
\end{center}

\subsection{Simulation}

Establishing the simulation relation follows the structure of the proof that $BBProt1$ refines $BBSpec1$.   $BBProt2$ essentially consists of an indexed collection of $BBProt1$ bulletin boards.  Each peer $j$ maintains a counter $p_j$ indicating its current bulletin board.   The bulletin board indexed by $p$ has $p < p_j$ in place of $pub_j$: Peer $j$ enters the publication phase for bulletin board $p$ once the counter $p_j$ has progressed beyond $p$.   It also has $p < c_j$ in place of $com_j$: Peer $j$ has commited to its share once the counter $c_j$ has progressed beyond $p$.  

Thus we obtain:
\begin{lemma}
$BBSpec2 \brefinedby BBProt2$ with respect to $\{ post, ack, publish \}$
\end{lemma}
\noindent {\bf Proof} (sketch)

We need to prove that if $J(s_A, s_C)$, and $s_C \stackrel{m_C}{\longrightarrow} s'_C$ then either $J(s_A, s'_C)$ ($m_c$ is matched by $skip$), or $\exists m_A, s'_A$ such that $s_A \stackrel{m_A}{\longrightarrow} s'_A$ and $J(s'_A, s'_C)$ ($m_c$ is matched by $m_A$).

The proof of each case for $m_C$ follows the same case in the proof of Lemma~\ref{lem:ref1}, where $p < p_j$ takes the place of $pub_j$.  We show two example cases: $c\_msg2a$ and $c\_dy2$ 

\noindent {\bf Case} $c\_msg2a$.  $Peer\; j \rightarrow DY : sig_{sk_{j}}(p_j,x)$.  
If $\# \{ k \mid 1 \le k \le t \land sig_{sk_{k}}(p_j,x) \in E \} = 2t-n-1$ and $\# \{ k \mid 1 \le k \le t \land sig_{sk_{k}}(p_j,x) \in E' \} = 2t-n$ then matched by $m_A = a\_msg1$ for $x, p_j$.  Otherwise matched by $skip$.  

\noindent {\bf Case} $c\_dy2$.  
If $x \in ITEM$ and $sig_{SSK}(p,x) \nin E$ and $sig_{SSK}(p,x) \in E'$, then this is matched by $a\_msg2$.  It remains to show that $x \in R_p$ and $sig_{SSK}(p,B) \in E_A \implies x \in B$.  The proof follows that of the same case in Lemma~\ref{lem:ref1}.

If $B_C \subseteq ITEM$ and $sig_{SSK}(p,B_C) \nin E$ and $sig_{SSK}(p,B_C) \in E'$, then this is matched by $a\_msg3$, with $Y = B_C $.  The proof that  (1) $E_A \inter PUBLISH_p = \{ \}$, 
(2) $C_p \subseteq Y$ and (3) $Y \subseteq R_p$, for $Y = B_C \}$ is entirely similar to this case in the proof of Lemma~\ref{lem:ref1}.

Otherwise matched by $skip$.

The other cases follow the same pattern.

This concludes the proof that $BBSpec2 \brefinedby BBProt2$ with respect to $\{post, ack, publish \}$.

\section{Accepting and Rejecting Posts} \label{sec:wbbcommit}

We now augment the Bulletin Board with an additional feature required for our use with Pr\^et \`a Voter: the ability to reject posts if they conflict with posts already received.  For example, different votes cannot be accepted on the same ballot, and audit requests cannot be accepted (even on different boards) after a vote has been cast.  

In particular, the bulletin board may refuse posts if they are inconsistent with previously accepted posts.  We express this by introducing an irreflexive symmetric binary relation $clash$ such that $clash(x,x')$ captures when two items $x$ and $x'$ should not both appear on the bulletin board.  For convenience we define $clashset(x) = \{ x' \mid clash(x,x') \}$ to be the set of all events that clash with $x$.

We will require that $x$ will be accepted if there is no $x'$ already received on any of the bulletin boards which clashes with $x$: in other words, that $clashset(x) \inter (\Union_p R_p) = \{ \}$.   

For example, in our context if $x$ is a vote on a ballot then $clashset(x)$ will be the set of audits and other votes on that ballot.  If $x$ is an audit on a ballot then $clashset(x)$ will be the set of all possible votes on that ballot.  If $c$ is a cancellation of a ballot then $clashset(x) = \emptyset$: a cancellation can always be added to the bulletin board.

\subsection{Specification}

The specification is obtained by simply strengthening the guard in $BBSpec2$ of the event $a\_msg1$ to include the non-clashing requirement.  All other events are identical to those in $BBSpec2$.  This yields the machine $BBSpec3$ as follows:

\begin{center} 
\framebox{
$\begin{array}{l}
\Bmachine BBSpec3 \\
\Bvariables  E_A, R_p, C_p \quad (p \in \nat)  \\ 
\qquad \vdots \\[0.5ex]
%
%\Binvariant  E_A \subseteq ITEM \union RECEIPT2 \union PUBLISH2 \\
%\phantom{\Binvariant}R_p \subseteq ITEM \quad (p \in \nat)  \\
%\phantom{\Binvariant}C_p \subseteq ITEM \quad (p \in \nat)  \\
%\Bevents  \\   
%\quad \mbox{init} \defs E_A := \{ \} \parallel \Parallel_{e \in \nat} (R_p := \{  \} \parallel C_p := \{ \}) \\[1ex]
%
%\quad  \mbox{post}(x) \defs \Bwhen x \in ITEM  \Bthen E_A := E_A \union \{ x \}  \Bend; \\[1ex]
%
%\quad r \longleftarrow \mbox{ack} \defs r :\in (E_A \inter RECEIPT2); \\[1ex]
%
%\quad  P \longleftarrow \mbox{publish} \defs P :\in (E_A \inter PUBLISH2); \\[1ex] 
%
\quad  \mbox{a\_msg1} \defs \\
\qquad \Bany x, p \\
\qquad \Bwhere x \in E_A \inter ITEM \land p \in \nat \land clashset(x) \inter (\Union_p R_p) = \{ \} \\
\qquad \Bthen R_p := R_p \union \{ x \}  \\
\qquad \Bend; \\[1ex]
%
%\quad \mbox{a\_msg2} \defs \\
%\qquad \Bany x, p \\
%\qquad \Bwhere x \in R_p \land \forall B . (sig_{SK_2}(p,B) \in E_A \implies sig_{SK1}(p,x) \in B) \\
%\qquad \Bthen E_A := E_A \union \{ sig_{SK_2}(p,x) \} \parallel C_p := C_p \union \{ x \} \\
%\qquad \Bend; \\[1ex]
%
%\quad \mbox{a\_msg3} \defs \\
%\qquad \Bany Y, p \\
%\qquad \Bwhere C_p \subseteq Y \subseteq R_p \\
%\qquad \Bthen E_A := E_A \union \{sig_{SK_2}(p,\{ sig_{SK1}(p,x) \mid x \in Y \} ) \} \\
%\qquad \Bend \\[1ex]
%
\qquad \vdots \\[0.5ex]
\Bend
\end{array}$
}
\end{center}

\subsection{Implementation}
We already have that $WBBProt2$ is already a refinement of $BBSpec2$.   Hence to obtain a refinement of $BBSpec3$ it is enough to strengthen the guards of the events in $WBBProt2$ matched by $a\_msg1$, to ensure that when they are enabled then so is $a\_msg1$.   In order to complete the refinement proof we also need to strengthen the invariant with clauses~\ref{invclash1} and \ref{invclash2} below.

In fact the only event matched by a\_msg1 in the proof of Lemma~\ref{lem:ref1} is c\_msg2a.  We will thus obtain machine $WBBProt3$ from $WBBProt2$ by strengthening c\_msg2a as follows:
\begin{center}
\framebox{
$\begin{array}{l}
\mbox{c\_msg2a$_j$} \defs \\
\quad \Bany x \\
\quad \Bwhere x\in I_{j,p_j} \land clashset(x) \inter \{ y \mid sk_{j}(p,y) \in \Union_p D_{j,p} \} = \{ \} \\
\quad \Bthen E := E \union \{ sig_{sk_j}(p_j,x) \} \parallel D_{j,p_j} := D_{j,p_j} \union \{ sig_{sk_j}(p_j,x) \}  \\
\quad \Bend;
\end{array}
$}
\end{center}

We also require two more clauses in the invariant of $WBBProt3$.  It is straightforward to establish that all of the events of $WBBProt3$ preserve these additional clauses:
\begin{eqnarray}
sig_{sk_j}(p,x) \in E & \iff & sig_{sk_j}(p,x) \in D_{j,p} \label{invclash1} \\
clash(x,x') \land sig_{sk_j}(p,x) \in D_{j,p} & \implies & sig_{sk_j}(p',x') \nin (\Union_p D_{j,p}) \label{invclash2} 
\end{eqnarray}

\subsection{Simulation}

With the exception of $a\_msg1$ and $c\_msg2a$, all events in $WBBSpec3$ and $WBBProt3$ are exactly the same as in $BBSpec2$ and $WBBProt2$, and so the refinements established previously remain valid.

We therefore only one new case to consider: $c\_msg2a$:

\noindent {\bf Case} $c\_msg2a$.  
If (1) $\# \{ k \le t \mid sig_{sk_{k}}(p_j,x) \in E \} = 2t-n-1$ and $sig_{sk_{j}}(p_j,x) \nin E$ and 
$clashset(x) \inter \{ y \mid sk_{j}(p,y) \in \Union_p D_{j,p} \} = \{ \}$
then this move will be matched by a\_msg1.  
Otherwise (2) $c\_msg2a$ is matched by $skip$ and we are done.  

Hence for (1) it remains to prove that the guard of a\_msg1 is enabled in this case, i.e. that $clashset(x) \inter (\Union_p R_p) = \{ \}$.
Since $J(s_A,s_C)$ this means that we must prove that in state $s_C$ there is no $x' \in clashset(x)$ such that $x' \in \Union_p R_p$, i.e. no $x'$ such that  
$\# \{ k \le t \mid sig_{sk_{k}}(p',x') \in E \} \ge 2t-n$.  

We establish this by contraction.  Assume there is some $x' $ such that $clash(x,x')$ and $\# \{ k \le t \mid sig_{sk_{k}}(p',x') \in E \} \ge 2t-n$ in state $s_C$.  This will also be the case in $s'_C$.  Also in state $s'_C$ we have $\# \{ k \le t \mid sig_{sk_{k}}(p_j,x) \in E \} = 2t-n$.  Hence by Corollary~\ref{cor:counting} there is some $k < t$ such that $sig_{sk_{k}}(p',x') \in E$ and $sig_{sk_{k}}(p_j,x) \in E$.   Hence by invariant~(\ref{invclash1}) we have  $sig_{sk_{k}}(p',x') \in D_{k'}$ and $sig_{sk_{k}}(p_j,x) \in D_{k,p_j}$.  This yields a contradiction with invariant~(\ref{invclash2}), since $clash(x,x')$.  

We thus conclude that the guard of $a\_msg1$ is enabled, and the refinement follows.

It follows that $BBSpec3 \brefinedby BBProt3$ with respect to $\{post, ack, publish \}$.

\subsection{Example: lower threshold allows acceptance of clashing posts}

\begin{example}
This example provides a second illustration as to why the threshold of honest peers is required to be greater than $2n/3$.  If it is not, then an adversary can arrange for receipts to be issued on clashing posts, as shown:
\def\svgwidth{0.75\linewidth}
\begin{center}
 %% Creator: Inkscape inkscape 0.48.2, www.inkscape.org
%% PDF/EPS/PS + LaTeX output extension by Johan Engelen, 2010
%% Accompanies image file 'clash.pdf' (pdf, eps, ps)
%%
%% To include the image in your LaTeX document, write
%%   \input{<filename>.pdf_tex}
%%  instead of
%%   \includegraphics{<filename>.pdf}
%% To scale the image, write
%%   \def\svgwidth{<desired width>}
%%   \input{<filename>.pdf_tex}
%%  instead of
%%   \includegraphics[width=<desired width>]{<filename>.pdf}
%%
%% Images with a different path to the parent latex file can
%% be accessed with the `import' package (which may need to be
%% installed) using
%%   \usepackage{import}
%% in the preamble, and then including the image with
%%   \import{<path to file>}{<filename>.pdf_tex}
%% Alternatively, one can specify
%%   \graphicspath{{<path to file>/}}
%% 
%% For more information, please see info/svg-inkscape on CTAN:
%%   http://tug.ctan.org/tex-archive/info/svg-inkscape
%%
\begingroup%
  \makeatletter%
  \providecommand\color[2][]{%
    \errmessage{(Inkscape) Color is used for the text in Inkscape, but the package 'color.sty' is not loaded}%
    \renewcommand\color[2][]{}%
  }%
  \providecommand\transparent[1]{%
    \errmessage{(Inkscape) Transparency is used (non-zero) for the text in Inkscape, but the package 'transparent.sty' is not loaded}%
    \renewcommand\transparent[1]{}%
  }%
  \providecommand\rotatebox[2]{#2}%
  \ifx\svgwidth\undefined%
    \setlength{\unitlength}{283.83904877bp}%
    \ifx\svgscale\undefined%
      \relax%
    \else%
      \setlength{\unitlength}{\unitlength * \real{\svgscale}}%
    \fi%
  \else%
    \setlength{\unitlength}{\svgwidth}%
  \fi%
  \global\let\svgwidth\undefined%
  \global\let\svgscale\undefined%
  \makeatother%
  \begin{picture}(1,1.13286089)%
    \put(0,0){\includegraphics[width=\unitlength]{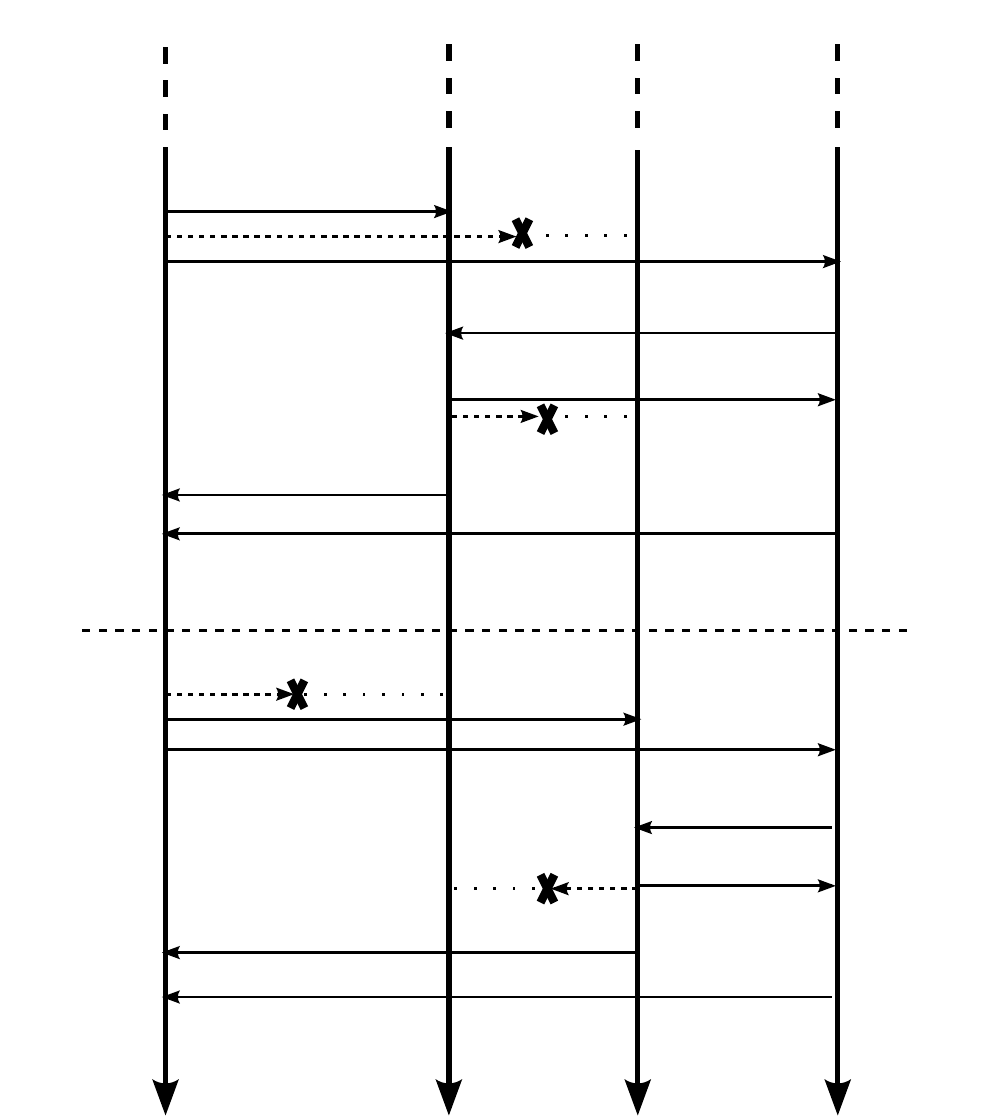}}%
    \put(0.13121746,1.10820453){\color[rgb]{0,0,0}\makebox(0,0)[lb]{\smash{User}}}%
    \put(0.41387264,1.10578867){\color[rgb]{0,0,0}\makebox(0,0)[lb]{\smash{Peer$1$}}}%
    \put(0.61116755,1.10538604){\color[rgb]{0,0,0}\makebox(0,0)[lb]{\smash{Peer$2$}}}%
    \put(0.80483866,1.10619132){\color[rgb]{0,0,0}\makebox(0,0)[lb]{\smash{E:Peer$3$}}}%
    \put(0.27884106,0.64911562){\color[rgb]{0,0,0}\makebox(0,0)[lb]{\smash{$sig_{ssk_1}(p,x)$}}}%
    \put(0.66316351,0.61227378){\color[rgb]{0,0,0}\makebox(0,0)[lb]{\smash{$sig_{ssk_3}(p,x)$}}}%
    \put(0.33722423,0.92935493){\color[rgb]{0,0,0}\makebox(0,0)[lb]{\smash{$x$}}}%
    \put(0.73161274,0.87801796){\color[rgb]{0,0,0}\makebox(0,0)[lb]{\smash{$x$}}}%
    \put(-0.00119692,0.55248132){\color[rgb]{0,0,0}\makebox(0,0)[lb]{\smash{$sig_{SSK}(p,x)$}}}%
    \put(0.00162158,0.5975773){\color[rgb]{0,0,0}\makebox(0,0)[lb]{\smash{construct}}}%
    \put(0.06443384,0.90962544){\color[rgb]{0,0,0}\makebox(0,0)[lb]{\smash{Post $x$}}}%
    \put(0.67061239,0.81379644){\color[rgb]{0,0,0}\makebox(0,0)[lb]{\smash{$sig_{sk_3}(p,x)$}}}%
    \put(0.4630501,0.7445419){\color[rgb]{0,0,0}\makebox(0,0)[lb]{\smash{$sig_{sk_1}(p,x)$}}}%
    \put(0.54035746,0.41759601){\color[rgb]{0,0,0}\makebox(0,0)[lb]{\smash{$y$}}}%
    \put(0.72557311,0.38618989){\color[rgb]{0,0,0}\makebox(0,0)[lb]{\smash{$y$}}}%
    \put(0.67524277,0.31854592){\color[rgb]{0,0,0}\makebox(0,0)[lb]{\smash{$sig_{sk_3}(p,y)$}}}%
    \put(0.65712388,0.25210993){\color[rgb]{0,0,0}\makebox(0,0)[lb]{\smash{$sig_{sk_2}(p,y)$}}}%
    \put(0.24320718,0.18607648){\color[rgb]{0,0,0}\makebox(0,0)[lb]{\smash{$sig_{ssk_2}(p,y)$}}}%
    \put(0.67926916,0.14057787){\color[rgb]{0,0,0}\makebox(0,0)[lb]{\smash{$sig_{ssk_3}(p,y)$}}}%
    \put(0.06483648,0.4188039){\color[rgb]{0,0,0}\makebox(0,0)[lb]{\smash{Post $y$}}}%
    \put(0.000011,0.14420163){\color[rgb]{0,0,0}\makebox(0,0)[lb]{\smash{construct}}}%
    \put(-0.00280749,0.09507926){\color[rgb]{0,0,0}\makebox(0,0)[lb]{\smash{$sig_{SSK}(p,y)$}}}%
  \end{picture}%
\endgroup%

\end{center}
In this attack two conflicting posts, $x$ and $y$, are both provided with receipts.  This is possible because no honest peer in involved in both:  Peer 1 and (dishonest) Peer 3 contribute to the receipt of $x$, and Peers 2 and 3 contribute to the receipt of $y$.  However, the WBB should only accept at most one of $x$ and $y$.  The attack is possible because Peer 3 provides shares towards the receipts of both $x$ and $y$, something no honest peer would do.  If the threshold of honest peers is greater than $2n/3$ then the same attack would require an honest peer to accept both $x$ and $y$, which the protocol prevents.
\end{example}

\section{Optimistic Commitment and Fallback} \label{sec:BB4}

The final element of the bulletin board to introduce is the optimistic protocol within the publication phase, and the use of signed hashes in publication of the bulletin board.  

\subsection*{Publication} 
When the time comes to publish, then the Post and Acknowledge protocol stops, and the peer begins the commit protocol which is used for the peers to obtain agreement on the bulletin board to publish.

Earlier models have used a round of message exchanges where peers circulate their database $D_j$, and another round where they circulate part-signed copies of their version of the bulletin board $t(D_j)$.    In the case where there is some disagreement on databases then peers can update their bulletin boards to include new items they have received.  

In practice we hope that in most cases the peers will agree on their local databases, and in this case they do not need to circulate them.  We therefore introduce an optimistic commit where they can simply circulate a partially signed hash of their bulletin board (together with the period $p$): if they agree on the hash then they combine to give a threshold signature, and any peer can publish the bulletin board with the signed hash.  If they do not agree then they can fall back to circulating their databases.

We therefore replace event $c\_msg6$ by two messages: one to circulate a part-signed hash of the bulletin board $sig_{ssk_j}(h(t(D_j)))$, and one to circulate the bulleting board $t(D_j)$ itself (since this cannot be retrieved from the hash).    The reason for separating these into two events is that we will eventually wish to schedule them separately: circulation of the hash will happen in the optimistic round, whereas publication of the board itself need not occur until there is agreement on the hash.

\begin{description}
\item[Optimistic commit protocol:]  This consists of two rounds:
\[
\begin{array}{lllll}
1. & P_i \rightarrow P_j  & : & sig_{sk_i}(p,h(B_{i,p})) & (\mbox{for each}\;i,j \in I, j \neq i) \\[0.5ex]
 \lefteqn{\qquad \quad \mbox{each $P_i$ checks the hashes from all peers agree}} \\[0.5ex] 
2. & P_i \rightarrow WBB & : & B_{i,p},\, sig_{ssk_i}(p,h(B_{i,p})) \qquad & (\mbox{for each}\;i \in I) 
\end{array}
\]
If there are not a threshold number of matching messages then the Fallback commit protocol is run:
\item[Fallback commit protocol:] This consists of a round of communications in which the peers exchange their  databases in order to make them consistent.
\[
\begin{array}{lllll}
1. & P_i \rightarrow P_j  & : & D_{i,p} \qquad & (\mbox{for each}\;i,j \in I, j \neq i)
\end{array}
\]
Peer $j$ receives $D_{i,p}$ from each of the other peers, up to some timeout.  For each $D$ received, Peer $j$ adds to $D_{j,p}$ any item $sig_{sk_k}(p,x) \in D$ that is not already in $D_{j,p}$.  

The peers then return to the optimistic commit protocol.
\end{description}

\subsection{Specification}

The specification $BBSpec4$ is similar to $BBSpec3$ except that the form of the published bulletin board is changed, so that the unsigned bulletin board is published together with a signed hash.    Events $a\_msg2$ and $a\_msg3$ are updated to reflect the change to the form in which bulletin boards are published, and $publish$ is also updated to output the new form of bulletin board.

\begin{eqnarray*}
RECEIPT4 & = & RECEIPT2 \\
PUBLISH4 & = & \{ sig_{SSK}(p,h(Y)) \mid p \in \nat \land Y \subseteq ITEM \} \\
PUBLISH4_p & = & \{ sig_{SSK}(p,h(Y)) \mid Y \subseteq ITEM \}
\end{eqnarray*}

\begin{center} 
\framebox{
$\begin{array}{l}
\Bmachine BBSpec4 \\
\Bvariables  E_A, R_p, C_p \quad (p \in \nat)  \\ 
\Binvariant  E_A \subseteq ITEM \union RECEIPT4 \union PUBLISH4 \\
\phantom{\Binvariant}R_p \subseteq ITEM \quad (p \in \nat)  \\
\phantom{\Binvariant}C_p \subseteq ITEM \quad (p \in \nat)  \\
\Bevents  \\   
\quad \mbox{init} \defs E_A := \{ \} \parallel \Parallel_{p \in \nat} (R_p := \{  \} \parallel C_p := \{ \}) \\[1ex]

\quad  \mbox{post}(x) \defs \Bwhen x \in ITEM  \Bthen E_A := E_A \union \{ x \}  \Bend; \\[1ex]

\quad r \longleftarrow \mbox{ack} \defs r :\in (E_A \inter RECEIPT4); \\[1ex]

\quad  P \longleftarrow \mbox{publish} \defs \\
\qquad \Bany Y, p \\
\qquad \Bwhere Y \subseteq ITEM \land sig_{SSK}(p,h(Y)) \in E_A \\
\qquad \Bthen P := (Y, sig_{SSK}(p,h(Y))) \\
\qquad \Bend;\\[1ex]

\quad  \mbox{a\_msg1} \defs \\
\qquad \Bany x \in E_A \inter ITEM \land p \in \nat \\
\qquad \Bthen R_p := R_p \union \{ x \}  \\
\qquad \Bend; \\[1ex]

\quad \mbox{a\_msg2} \defs \\
\qquad \Bany x, p \\
\qquad \Bwhere x \in R_p \land (sig_{SSK}(p,h(B)) \in E_A \implies x \in B) \\
\qquad \Bthen E_A := E_A \union \{ sig_{SSK}(p,x) \} \parallel C_p := C_p \union \{ x \} \\
\qquad \Bend; \\[1ex]

\quad \mbox{a\_msg3} \defs \\
\qquad \Bany Y, p \\
\qquad \Bwhere C_p \subseteq Y \subseteq R_p \land E_A \inter PUBLISH_4 = \{ \} \\
\qquad \Bthen E_A := E_A \union \{ sig_{SSK}(p,h(Y)) \} \\
\qquad \Bend \\[1ex]

\Bend
\end{array}$
}
\end{center}

\subsection{Implementation}

Shared signatures on the bulletin board are now on its hash value.  The effects of this change is highlighted in the invariant below.  The remaining clauses remain the same as in $BBProt3$.

\begin{center} 
\framebox{
$\begin{array}{l}
\Bmachine BBProt4 \\[1ex]
\Brefines BBSpec4 \\[1ex]
\Bvariables  E,\; I_{j,p},\; D_{j,p},\; H_{j,p},\; p_j \;\; (1 \le j \le t) \\[2ex]
\Binvariant  \\
\mbox{/* Types */} \\[0.5ex]
\qquad p_j \in \nat \\[0.5ex]
\qquad E \subseteq MESSAGE \\[0.5ex]
%\phantom{\Binvariant} \forall j \;.\; \\[0.5ex]
\qquad I_{j,p} \subseteq ITEM \\[0.5ex]
\qquad
D_{j,p} \subseteq \{ sig_{sk_k}(p,x) \mid x \in ITEM \} \\[0.5ex]
\qquad
H_{j,p} \subseteq \{ sig_{sk_k}(p,h(B)) \mid B \subseteq ITEM \} \\[0.5ex]
\mbox{/* Key invariant properties */} \\[0.5ex]
\qquad
 k \le t \land k \in c[p,x] \land k \in s[p,B]  \Rightarrow x \in B \hspace*{2.4in} \\[0.5ex]
\qquad
 sig_{ssk_j}(p,x) \in E  \implies  \# d_j[p,x] \ge t \\[0.5ex] 
\qquad
 sig_{SSK}(p,x) \in E  \implies  \# c[p,x] \ge t \\[0.5ex] 
\qquad
\fbox{$sig_{ssk_j}(p,h(B)) \in E \implies B \subseteq t(D_{j,p})$} \\[0.5ex]
\qquad
\fbox{$sig_{SSK}(p,h(B)) \in E  \implies  \# s[p,B] \ge t$} \\[0.5ex] 
\qquad
 D_{j,p}  \subseteq  E \\[0.5ex]
 \qquad
 k \le t \land k \in s[p,B]  \Rightarrow c_k > p \\[0.5ex]
\qquad
 k \le t \land k \in s[p,B_1] \land B_1 \neq B_2 \Rightarrow k \nin s[p,B_2] \\[0.5ex]

{\mbox{ /* linking invariant */}} \\[0.5ex]
\qquad
R_p  =  \{ x \in ITEM \mid \# \{ k \mid 1 \le k \le t \land sig_{sk_{k}}(p,x) \in E \} \ge 2t-n \}  \\[0.5ex]
\qquad
C_p  =  \{ x \in ITEM \mid sig_{SSK}(p,x) \in E \} \\[0.5ex]
\qquad
E_A =  E \inter (ITEM \union RECEIPT4 \union PUBLISH4) \\[0.5ex]
\mbox{where:} \\
\qquad
d_j[p,x] = \{ k \mid sig_{sk_{k}}(p,x) \in D_j \}  \qquad \mbox{shares of part sigs on $x$ received by Peer $j$}\\[0.5ex]
\qquad
c[p,x]  \hspace*{1.2mm} =  \{ k \mid sig_{ssk_{k}}(p,x) \in E \}  \qquad  \mbox{peers which have (part)signed the receipt on $x$} \\[0.5ex]
\qquad
\fbox{$s[p,B] \hspace*{0.5mm} =  \{ k \mid sig_{ssk_{k}}(p,h(B)) \in E \}$} \qquad  \mbox{peers which have part-signed bulletin board $B$} 
\end{array}$
}
\end{center}

\subsubsection*{External events}

External events are very similar in $BBProt4$.  The event $publish$ is adjusted to reflect the new form of publication, but external evetns are otherwise the same as in $BBProt3$.

\begin{center} 
\framebox{
$\begin{array}{l}
\Bevents  \\   
\quad \mbox{init} \defs E := \{ sk_{k} \mid k > t \} \union \{ ssk_{k} \mid k > t \}  \parallel \\
\phantom{\quad \mbox{init} \defs} \Parallel_{j,n} (I_{j,n} := \{ \} \parallel D_{j,n} := \{ \} \parallel  H_{j,n} := \{ \} \parallel p_j := 0 \parallel c_j := 0); \\[2ex]

\quad  \mbox{post}(x) \defs \Bwhen x \in ITEM \Bthen E := E \union \{ x \}  \Bend; \\[1.5ex]

\quad r \longleftarrow \mbox{ack} \defs 
r :\in (E \inter RECEIPT4); \\[1ex]

\quad  P \longleftarrow \mbox{publish} \defs \\
\qquad \Bany Y, p \\
\qquad \Bwhere Y \subseteq ITEM \land Y \in E \land sig_{SSK}(p,h(Y)) \in E \\
\qquad \Bthen P := (Y, sig_{SSK}(p,h(Y))) \\
\qquad \Bend;\\[1ex]

\end{array}$
}
\end{center}

\subsubsection*{Posting and acknowledgement protocol}
Posting and acknowledgement is identical to $BBProt3$, and so we do not reproduce the events here.

\subsubsection*{Commit and publish protocol}

Agreeing and publishing the bulletin board now has two additional events: sending the bulletin board explicitly in $c\_msg7$, and circulating signed hashes in the optimistic phase $c\_msg8$.  Note that $c\_msg6$ now provides a partially signed hash rather than a partially signed bulletin board.  
\begin{center} 
\framebox{
$\begin{array}{l}
\quad \mbox{c\_msg4$_j$} \defs \qquad \mbox{/* start commit protocol */} \\
\qquad \Bbegin \\
\qquad p_j := p_j + 1 \\
\qquad \Bend; \\[1ex]

\quad \mbox{c\_msg5a$_j$} \defs \qquad \mbox{/* send database */} \\
\qquad \Bany p < p_j \\
\qquad \Bthen E := E \union  \{ D_{j,p} \} \\
\qquad \Bend; \\[1ex]

\quad \mbox{c\_msg5b$_j$} \defs \qquad \mbox{/* receive database */} \\
\qquad \Bany D, p \\
\qquad \Bwhere D\in E \land D\subseteq SIG1_p \land p < p_j \\
\qquad \Bthen D_{j,p} := D_{j,p} \union D \\
\qquad \Bend; \\[1ex]

\quad \mbox{c\_msg6$_j$} \defs \qquad \mbox{/* provide partially signed hash */} \\
\qquad \Bwhen c_j < p_j \land \# \{ k \mid sig_{sk_k}(c_j,h(t(D_{j,c_j})) \in H_{j,c_j} \} \ge t \\
\qquad \Bthen E := E \union \{ sig_{ssk_j}(c_j,h(t(D_{j,c_j}))) \} \parallel c_j := c_j + 1 \\
\qquad \Bend; \\[1ex]

\quad \mbox{c\_msg7$_j$} \defs \qquad \mbox{/* send bulletin board */} \\
\qquad \Bany p < p_j \\
\qquad \Bthen E := E \union  \{ t(D_{j,p}) \} \\
\qquad \Bend; \\[1ex]

\quad \mbox{c\_msg8a$_j$} \defs \qquad \mbox{/* send signed hash */} \\
\qquad \Bany p < p_j \\
\qquad \Bthen E := E \union \{ sig_{sk_j}(p,h(t(D_{j,p}))) \} \\
\qquad \Bend; \\[1ex]

\quad \mbox{c\_msg8b$_j$} \defs \qquad \mbox{/* receive signed hash */} \\
\qquad \Bany p < p_j \land k \le n \land B \subseteq ITEM \land sig_{sk_k}(p,h(B)) \in E \\
\qquad \Bthen H_{j,p} := H_{j,p} \union  \{  sig_{sk_k}(p,h(B)) \} \\
\qquad \Bend; \\[1ex]

\end{array}$
}
\end{center}

\subsubsection*{Dolev-Yao environment}
The adversary has the same moves as before, with one additional one: hashing.  Any message can be hashed.
\begin{center} 
\framebox{
\begin{minipage}{0.5\textwidth}
$\begin{array}{l}
 \mbox{c\_dy8} \defs \quad \mbox{/* hashing */}\\
\quad \Bany m \\
\quad \Bwhere m \in E \\
\quad \Bthen E := E \union \{ h(m) \} \\
\quad \Bend 
\end{array}$
\end{minipage}
}
\end{center}

\subsection{Simulation}

The proof of simulation follows exactly the pattern of previous proofs.  In particular:
\begin{itemize}
\item The external events $post$, $ack$ and $publish$ of $BBSpec4$ are refined by their counterparts in $BBSpec4$.   
\item Event $a\_msg1$ is refined by the appropriate occurrence of $c\_msg2a$.
\item Events $a\_msg2$ and $a\_msg3$ are refined by $c\_dy2$ combining the signature shares appropriate to each case.
\item All other concrete events  refine $skip$.   In particular, the new events of the optimistic protocol $c\_msg7$ and $c\_msg8$ refine $skip$.
\end{itemize}

This concludes the proof that $BBSpec4 \brefinedby BBProt4$ with respect to $\{post, ack, publish \}$, establishing the correctness of the bulletin board protocol $BBProt4$ against the specification $BBSpec4$.

\section{Liveness} \label{sec:liveness}

The Dolev-Yao threat model does not allow the protocol to provide any liveness guarantees.  All communications between the parties involved in the protocols can be blocked, preventing protocols from completing.  The threat model is appropriate for consideration of safety properties, but is too strong for analysis of liveness.  We require some assumptions about the communications between the protocol participants, as well as their honesty, in order to consider liveness.

We are primarily concerned with liveness for the publication of the bulletin board at the end of each period.  To reason about liveness we assume that communication between peers is reliable, but that some of the peers may not follow the protocol, either because they are dishonest, or because they have failed.   

\subsection{All honest peers}

We consider the case where all bulletin board peers are honest and follow the protocol.  This scenario includes the case where users may be dishonest, sending different information to different peers, or not involving peers in some posting rounds.  It also allows for the possibility where peers have not synchronised perfectly on the end of the period, so some posted items may be received in different periods for different peers, and hence their local records of the bulletin boards will not match.

Liveness can be shown for the commit and publish protocol.  Different peers may begin that protocol with different databases $D_i$ from the postings in the period.  The optimistic protocol may complete if enough of them agree on $t(D_i)$, the contents of the bulletin board.  However, it might not complete if the peers have sufficiently different records of what the bulletin board should contain.  In that case all peers execute the fallback protocol, and so communicate their databases reliably to all other peers.  This results in all peers ending up with the same database record of posted items (i.e. the union of all their databases), and thus the second execution of the optimistic protocol will succeed in generating a threshold signature on the bulletin board.

\subsection{A threshold of honest peers, and honest users}

We now consider the case where some peers are not in communication for the commit and publish protocol.
However we assume a threshold are behaving correctly and communicating with one another.  

If every post of an item during the posting phase involved a threshold of (not necessarily the same) peers and obtained a receipt, then a single round of the fallback phase will ensure that all posted items are now obtained by all of the live peers.  They are all sharing their evidence, and for each post there is at least one honest peer who is live in the commit and publish phase and also participated in the receipt of that item.  This peer will provide the evidence of receipt to the other peers in the fallback round.  Hence they will all agree on all posted items in the subsequent optimistic round.  In this case again only one fallback round is required before agreement was obtained.

\subsection{A threshold of honest peers}

We now consider the more general case, where only a threshold set of peers are honest and connected during the publication phase.   All peers will fix on a database $D_i$ when they enter the publication phase: honest peers will use the $D_i$ corresponding to the item posts they have received in the period, and we allow that dishonest peers will select any arbitrary $D_i$ within their capability.    We assume that peers will not change their database once it has been fixed, and will not send different databases to different peers, since this form of dishonesty would be easy to detect in practice.  If it is detected then the dishonest peer would be removed and the protocol re-run (corresponding to a complete failure of that peer).  Thus the only failure we need consider for peers outside the threshold set is failure to communicate, known as a  stopping failure.

With this form of failure the {\em Floodset} agreement algorithm \cite[6.2]{DBLP:books/mk/Lynch96} will lead to all honest peers agreeing on a database, and hence a bulletin board, within a maximum of $n-t+1$ rounds.  Each round of the Floodset algorithm is essentially the fallback protocol, with the optimistic protocol checking after each round whether there is a consensus.   In the context of our commit and publish protocol the peers are looking for agreement on the union of their values rather than on one particular value, and so take their agreed value to be that set.

\section{Discussion} \label{sec:conc}

\subsection{Summary}

In this paper we have presented a distributed protocol for running a bulletin board using a number of peers, which can tolerate a number of them failing, in the context of a threat model which has the communication between the peers controlled by a Dolev-Yao adversary, who also is able to control some of the peers.  This provides robustness and distributed trust: the bulletin board can tolerate some peers failing, and we require that a threshold of peers should be honest but no individual peer is required to be trusted.  Provided a threshold of the peers behave according to the protocol, the key properties demanded of the bulletin board hold.  In particular, only items posted to the bulletin board will be posted on it, any item whose receipt is acknowledged by the bulletin board must be posted on it, and the bulletin board will not accept conflicting items.  The bulletin board protocol has also been shown to be live when a threshold of honest peers all  communicate without interference with each other, even if some dishonest peers attempt to disrupt progress.

The development of this modelling and verification approach for this kind of protocol is also one of the contributions of this paper.  Correctness has been established formally using the Event-B framework, using simulation to show that the protocol is a refinement of an idealised bulletin board which has the desired properties.  The model included a Dolev-Yao attacker and the description of the protocol steps followed by the peers.  Carrying out the proof identified some nondeterminism inherent in the protocol and enabled us to include it in the idealisation to document the possible behaviour of the implementation.   In particular, an adversary can create a situation where he controls whether or not an unreceipted item appears on the bulletin board, and so this is reflected as nondeterminism at the abstract level.  In the context of the vVote system this will not be an issue in practice since the voting ceremony requires that any unreceipted items should be cancelled.  Hence the nondeterminism will not affect the tallying of the election: either the cancellation appears alongside a vote, or it appears without the vote.

\subsection{Related work}

Other proposals for bulletin board implementations using a set of peers are given by Krummenacher \cite{krummenacher}, by Peters \cite{peters05:e-vote}, and in the STAR-VOTE system \cite{starvote-jets}.

\subsubsection*{Krummenacher's Bulletin Board}
Krummenacher focuses on a peered bulletin board that guarantees the correctness of its history and the authenticity of the messages.  His proposal is motivated by the desire to provide a distributed version of Heather and Lundin's append-only web bulletin board \cite{DBLP:conf/ifip1-7/HeatherL08}.  The protocol is designed essentially for robustness, and is considered in the context where up to $k$ out of $n$ peers may fail.   A particular number of peers ($k+1$) must accept an item for it to be allowed onto the bulletin board.  Peers hold their own versions of the history of items posted, and so their histories would need to be combined in order to obtain the global bulletin board.   A similarity with our approach is the need for peers to confirm that other peers have received an item before providing their own response.  However, peers use a locking protocol when they seek confirmation from other peers, so the approach does not scale up well as $k$ gets larger relative to $n$.   

The most significant difference with the approach of this paper is the threat model: Krummenacher considers the protocol in the context of communication failures and peer failures, but does not consider an active adversary or corrupt peers who might deliberately introduce invalid messages (or accept clashing items).   Formal modelling and analysis would help to clarify the adversarial context and identify whether the protocol does indeed guarantee correct behaviour within that threat model.  Another difference is that Krummenacher's bulletin board is not concerned with preventing clashing items from being posted.  This might be addressed by setting the threshold $k$ to be sufficiently high and requiring that individual peers do not accept items that clash.  If peers can be dishonest then we may require the threshold of $k > 2n/3$, but this threshold does not work well with the locking protocol used in posting.   Finally, we also observe that the protocol follows the approach of \cite{DBLP:conf/ifip1-7/HeatherL08} in using timestamps to ensure that the bulletin board is relatively recent, and hence that no commit round is required.  Instead the peers are always able to provide their current version of the bulletin board.  This approach gives rise to challenges in implementation, notably that a single view of the `official' bulletin board, as would be required in an election context, would need to be constantly refreshed by the bulletin board peers.  This would be a substantial overhead, and furthermore its security implications are not well understood.    For all these reasons Krummenacher's implementation is not suitable as it stands for our requirements. 

\subsubsection*{Peters' Bulletin Board}
Peters \cite{peters05:e-vote} considers several approachs, and proposes and implements a bulletin board which uses a secure agreement protocol \cite{DBLP:conf/ccs/Reiter94}  on top of a group membership protocol \cite{DBLP:journals/tse/Reiter96}.   Items are posted by a client to a single peer, which then communicates with the others, obtains confirmation of receipt from a threshold of them, circulates that confirmation back to the peers, and returns a receipt to the client.  This approach is similar to our posting protocol, where peers require confirmation of receipt from a threshold of other peers, before returning their share of the signature on the receipt.  The system requires the same threshold as we do: that strictly more than $2/3$ of the peers behave honestly.  Further, each honest peer can serve the complete bulletin board on request.  They achieve this by means of a round of communication collecting signature shares on the bulletin board after each item is posted, similar to our optimistic protocol for the end-of-period publication.  In practice this might carry an overhead, both in obtaining the agreement and in providing the bulletin board, and in our context it is not necessary.  However, it would be perfectly possibly to run the bulletin board and only carry this out at the end of the period.

Similar to our approach, peers can also reject posts that clash with previous posts (such as a second vote on the same ballot form), and the threshold ensures that the collective bulletin board will not accept posts that clash.  Dishonest peers are handled by use of the group membership protocol: a current group of participating peers is maintained by all honest peers, and dishonest peers once detected are removed from the group.  Peers can also be readmitted to the group, in which case they need to bring themselves up to date on the state of the bulletin board.    

A key difference with our approach is the use of the group membership protocol to dynamically change the set of `live' peers.  Peters provides excellent formal descriptions of the protocols with sufficient detail to enable code production, and also gives arguments of their correctness in this setting.  However, there is no formal verification of the collection of protocols operating together.  The possible interactions between them are quite subtle and require careful handling, for example how reconfiguration of the group might interact with the posting of items, or how a client may need to switch from an ejected peer to an honest one.  Some dishonest peer behaviour might not trigger removal from the group, but might still interfere with the protocols, and this possibility requires careful analysis.    A second key difference is the way the (honest) peers need to keep a record of the up to date bulletin board at all times.  Although honest peers in our system will also have a record of the bulletin board if they are connected and participate in the posting of items, it is not a requirement, and peers are not relied on for it.   A final key difference is in the threat model, which allows dishonest peers but considers the network itself to be reliable (as we do for liveness), so honest peers can always communicate with each other.   A peer failing to communicate is treated as dishonest.  This will trigger a reconfiguration of the group and makes the protocol more sensitive to minor communication failures.

\subsubsection*{STAR-Vote}
The use of the bulletin board within STAR-Vote is close to ours: it collects votes during the election, and publishes only at the end.  The voting terminals are networked and play the role of bulletin board peers: they collect the votes as they are cast, and track which ones are to be counted.  The voter retains a paper receipt as evidence of their vote, which does not reveal anything of how they voted.  The system is designed to tolerate faulty peers.  At the end of the election the voting terminals agree on the votes that have been received, and publish them in encrypted form on the web so that voters can check them against their receipts.  The electronic record is also checked against the paper copies of the votes retained by the system.  The set of votes is signed by the election authorities to prevent subsequent manipulation.  Following the approach of VoteBox\cite{VoteBox}, the Voting Terminals maintain a global audit log during the election using a hash chain, so received votes are committed to in real time, and past events cannot be tampered with by a subset of malicious machines.

The main difference with our approach is that STAR-Vote is designed for use in a single polling station.  This gives a different threat model, in particular the threats are considered to come from malicious devices rather than the underlying network.  There is no geographic separation between casting a vote and having it received by the bulletin board peers, and the local network is assumed to provide reliable communication.  STAR-Vote therefore does not need to address the challenge of posting items to a remote bulletin board, which we have had to address by having the posting protocol generate a cryptographically signed receipt to provide the evidence that the item has been received.  Currently STAR-Vote does not provide signatures on the receipts that voters retain, though this is considered as a possibility in the context of mitigation against a ``defaming'' attack where voters present falsified evidence against the bulletin board.  STAR-Vote also does not go into detail about how the bulletin board information is collated from the peers, in particular what happens when some voting terminals but not others claim to have received a vote.  The emphasis is on detection of incorrect behaviour rather than its automatic correction.  If discrepancies are identified, then the approach would be to resolve them forensically, by checking audit logs, memory dumps, and other relevant records.

\subsubsection*{Byzantine Agreement Protocols}
The general problem of achieving generalised agreement across components where some might fail in adverse ways is known as the {\em Byzantine Agreement} problem \cite{DBLP:journals/toplas/LamportSP82}, and there is an extensive literature on approaches to the problem \cite{DBLP:books/mk/Lynch96}.  Such protocols require correct behaviour in strictly more than $2/3$ of the peers, the same requirement as we have on our bulletin board peers.   However, Byzantine Agreement protocols are not really suitable for items being posted.  These protocols are typically synchronous and proceed in rounds, which would be too inefficient for receiving large quantities of votes: too many rounds and too much synchronisation overhead would be required to process each vote if we wish the peers to agree on the receipt of every vote.  Furthermore, not all peers would necessarily be aware when a protocol run is starting, since they may not receive the initial item.  Instead we have provided a protocol for the peers simply to send messages to each other and to respond to messages received in an asynchronous fashion.  This means that the peers do not all need to agree on each vote.  Our threat model is also different to the typical threat model for Byzantine agreement protocols: ours allows honest peers to be excluded from the acceptance of some items to the bulletin board, without being considered as dishonest, whereas Byzantine agreement protocols consider any non-participating peer as failing.   

We are closer to the problem of Byzantine agreement in the commit and publish phase, since this is where the peers seek consensus to agree on a bulletin board to publish.  Our optimistic and fallback protocols are indeed close to the Floodset protocol \cite{DBLP:books/mk/Lynch96}, a basic agreement protocol.  Even in this case we do not require the full power of Byzantine agreement protocols: the use of signatures minimises the ability of dishonest peers to introduce additional confusing information to disrupt the protocol run, and we can limit the adversarial behaviour simply to peers ceasing to communicate.

\subsection{Implementation level considerations}

The concrete model above has been analysed for correctness, and shown to be correct with respect to the abstract model.  However, even the concrete model is nondeterministic in the order in which events should be scheduled.  This is deliberate, since it means that interactions between protocols are addressed in the analysis, but in practice we will want an efficient implementation and so will schedule events in a particular way, to avoid expensive computations such as the fallback protocol unless they are necessary, or for other pragmatic reasons.  For example, the implementation we have developed for the vVote project requires that all peers should agree on the hash of the bulletin board in order to provide their signature share, although only a threshold of hashes on the database is sufficient for correctness.  We do this because it is still helpful to know if possible that all peers have the same database, to provide reassurance that the protocol is working properly.   However, this implementation is consistent with the concrete model, in which peers can start the fallback protocol at any time.  As long as the implementation performs events in accordance with the concrete protocol, it will provide behaviour that is correct with respect to the abstract specification.

Once published, a web bulletin board will need provide voters with the facilities to confirm their vote is correctly recorded within the signed bulletin board, and to be able to obtain the full contents of the signed bulletin board so that the subsequent processing can be checked.  In order to ensure that the board cannot be later replaced with a different signed bulletin board, the signed hash of the board will also be published at the end of the period using an out-of-band broadcast channel.  For the planned use of the system in Victoria in November 2014, each period will be one day, and the signed hash will be published the following day in the newspaper.  Voters can then check the bulletin board on the web against that published hash.

There are different ways to make the contents of the bulletin board available.  To be consistent with the commit and publish protocol, all that is needed is the peers are able to produce shares on a threshold signature of what is produced.  It may simply make the entire bulletin board available for download, and then the voter will check its signature and that their vote is included (and not cancelled) within it.  Alternatively it may make use of a structure such as a hash tree \cite{DBLP:journals/corr/abs-0908-4116} which provides signature authentication that an item is included on the bulletin board without the need to download the entire board.  In our case we need to check not only that a vote is present, but also that there is also no cancellation present.  In practice this may be done by also publishing and signing the entire set of cancellations, since this is likely to be small compared to the set of votes cast.  Checking the presence of a vote then involves confirming both that a vote is present in the hash tree and that it has not been cancelled.   This is the subject of ongoing research.

\subsubsection*{Acknowledgements}

We are grateful to James Heather, Peter Ryan, Vanessa Teague and Douglas Wikstr\"om for useful discussions on Bulletin Board design and approaches to verification.  We are also grateful to Gavin Lowe and Joshua Guttman for detailed discussions on the formal modelling and verification approach, and to Thierry Lecomte, Helen Treharne, John Derrick and Graeme Smith for discussion and advice on the B modelling and refinement.  Thanks also to Olivier Pereira and Dan Wallach for clarifying aspects of STAR-Vote.  This work was supported by the EPSRC Trustworthy Voting Systems project  EP/G025797/1.

\bibliographystyle{alpha}
\bibliography{Bib,e-vote}

\end{document}